\documentclass[lettersize,journal]{IEEEtran}
\usepackage{amsmath,amsfonts}
\usepackage{algorithm}
\usepackage[noend]{algpseudocode}
\usepackage{array}
\usepackage{subcaption}
\usepackage{booktabs}
\usepackage{textcomp}
\usepackage{stfloats}
\usepackage{url}
\usepackage{hyperref}
\usepackage{verbatim}
\usepackage{graphicx}
\usepackage{bm}
\usepackage{cite}
\usepackage{orcidlink}
\begin{document}

\title{ERX: A Fast Real-Time Anomaly Detection Algorithm for Hyperspectral Line Scanning}

\author{Samuel~Garske \orcidlink{0000-0002-0622-8255},~\IEEEmembership{Graduate Student Member,~IEEE,}
        Bradley~Evans \orcidlink{0000-0001-6675-3118},
        Christopher~Artlett \orcidlink{0000-0001-9809-7098},
        and~KC~Wong \orcidlink{0000-0002-4977-5611}% <-this % stops a space

%\thanks{Manuscript received Month DD, YYYY; revised Month DD, YYYY.}%
\thanks{This work has been submitted to the IEEE for possible publication. Copyright may be transferred without notice, after which this version may no longer be accessible. This work was supported by the Australian Research Council through the Industrial Transformation Training Centre grant IC170100023 that funds the Australian Research Council (ARC) Training Centre for CubeSats, UAVs \& Their Applications (CUAVA).}% <-this % stops a space
\thanks{Samuel Garske and KC Wong are with the School of Aerospace, Mechanical, and Mechatronic Engineering, The University of Sydney, NSW 2006, Australia (e-mail: sam.garske@sydney.edu.au). Bradley Evans is with the School of Environmental and Rural Science, The University of New England, Armidale, NSW 2350, Australia. Christopher Artlett is with the Defence Science and Technology Group, Eveleigh, NSW 2015, Australia. Samuel Garske, Bradley Evans, and KC Wong are also with CUAVA.}}% <-this % stops a space

\maketitle

\begin{abstract}
Detecting unexpected objects (anomalies) in real time has great potential for monitoring, managing, and protecting the environment. Hyperspectral line-scan cameras are a low-cost solution that enhance confidence in anomaly detection over RGB and multispectral imagery. However, existing line-scan algorithms are too slow when using small computers (e.g. those onboard a drone or small satellite), do not adapt to changing scenery, or lack robustness against geometric distortions. This paper introduces the Exponentially moving RX algorithm (ERX) to address these issues, and compares it with four existing RX-based anomaly detection methods for hyperspectral line scanning. Three large and more complex datasets are also introduced to better assess the practical challenges when using line-scan cameras (two hyperspectral and one multispectral). ERX is evaluated using a Jetson Xavier NX edge computing module (6-core CPU, 8GB RAM, 20W power draw), achieving the best combination of speed and detection performance. ERX was 9 times faster than the next-best algorithm on the dataset with the highest number of bands (108 band), with an average speed of 561 lines per second on the Jetson. It achieved a 29.3\% AUC improvement over the next-best algorithm on the most challenging dataset, while showing greater adaptability through consistently high AUC scores regardless of the camera's starting location. ERX performed robustly across all datasets, achieving an AUC of 0.941 on a drone-collected hyperspectral line scan dataset without geometric corrections (a 16.9\% improvement over existing algorithms). This work enables future research on the detection of anomalous objects in real time, adaptive and automatic threshold selection, and real-time field tests. The datasets and the Python code are openly available at \href{https://github.com/WiseGamgee/HyperAD}{https://github.com/WiseGamgee/HyperAD}, promoting accessibility and future work.
\end{abstract}


% \fi
\IEEEpeerreviewmaketitle

\section{Introduction}

\IEEEPARstart{A}{nomaly} detection is an extremely useful task for finding unexpected objects in an image. It is an unsupervised learning task where no prior information or annotated examples of the anomalies are needed \cite{chang2002anomaly}. Detecting volcanic hotspots \cite{corradino2022volcanicAD}, identifying defects in manufactured products \cite{bergmann2019mvtec}, and finding medical markers for the diagnosis of disease \cite{schlegl2017medicalAD} are examples of image-based anomaly detection. 

\begin{figure}
    \centering
    \begin{minipage}{\linewidth}
        \centering
        \includegraphics[width=\linewidth]{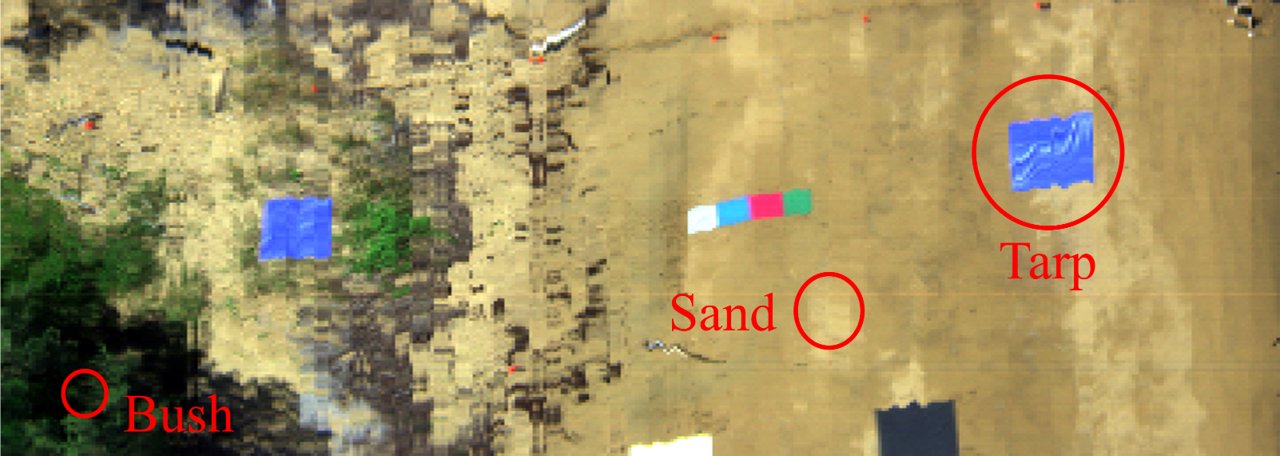}
        \caption*{} % No caption
    \end{minipage}

    \begin{minipage}{\linewidth}
        \centering
        \includegraphics[width=\linewidth]{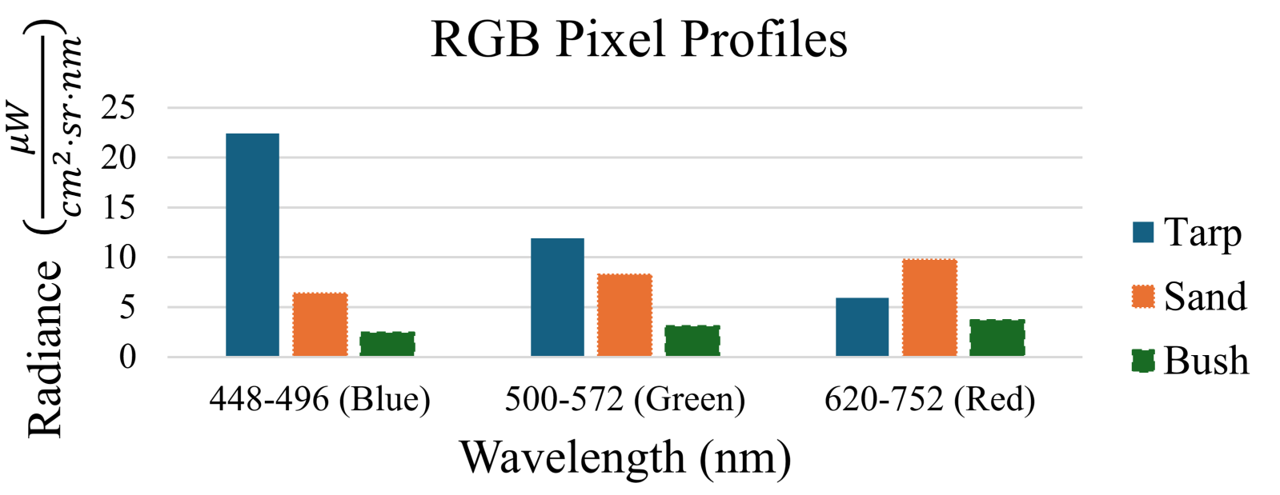}
        \caption*{} % No caption
    \end{minipage}

    \begin{minipage}{\linewidth}
        \centering
        \includegraphics[width=\linewidth]{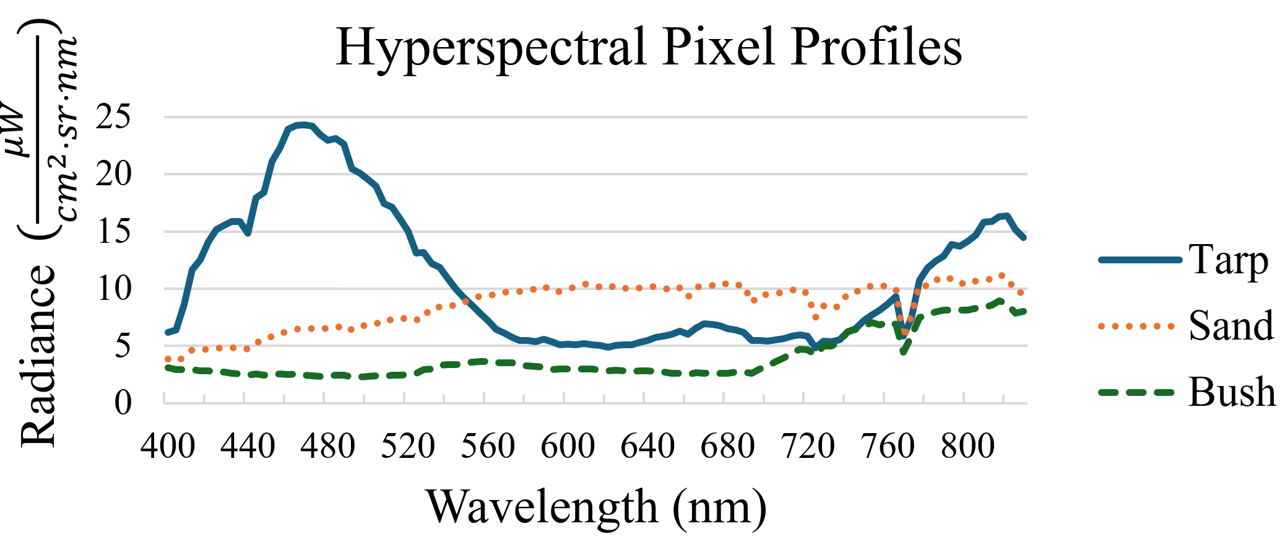}
        \caption*{} % No caption
    \end{minipage}
    \caption{Comparison of RGB vs HSI pixel profiles for various objects. The top image shows a section of the beach dataset from section \ref{section_datasets}, with pixels extracted from a tarp, sand, and bush (vegetation). Hyperspectral pixels (bottom) provide a detailed and continuous spectral profile via narrower bands, compared to the the discrete RGB counterparts (middle). The RGB bands use the average radiance across the defined wavelength range.}
    \label{fig:RGBvsHSI}
\end{figure}

Detecting anomalies in real time is essential for applications requiring rapid insights and fast decision making. This enables immediate anomaly mapping when surveying a region of interest, and onboard system responses to detected anomalies. Onboard detection can also reduce data transmission requirements by sending only key insights rather than raw, multi-channel image data. These capabilities are particularly valuable in hyperspectral imaging for remote sensing, a rapidly growing area of research used to monitor, manage, and protect the environment \cite{asner2017forest, farmonov2023desis, pei2023prisma, papp2021drone, green2020emit, lehmann2023gloria}. Hyperspectral imaging, also known as imaging spectroscopy, provides greater confidence than RGB or multispectral imaging when detecting anomalies. This is because the pixels in a hyperspectral image (HSI) have more detailed spectral profiles that uniquely represent the material present \cite{matteoli2010HADreview}, as shown in Figure \ref{fig:RGBvsHSI}. The greater depth of spectral information can make it easier to separate anomalies from the background \cite{chang2002anomaly}.

\begin{figure}[h!]
    \centering
    \includegraphics[width=\linewidth]{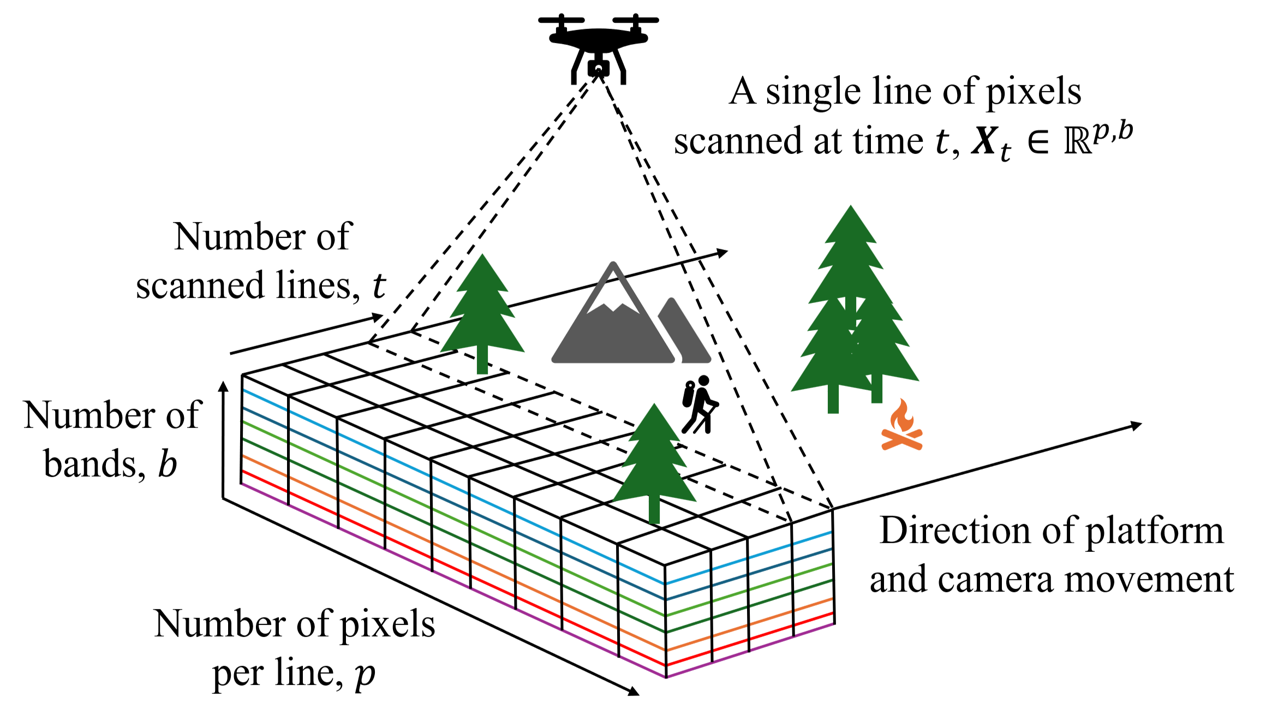}
    \caption{A line-scan camera capturing one line of pixels at a time as it's platform (e.g., a drone) moves over the area of interest. These lines form a hyperspectral image, or datacube, with the depth dimension $b$ representing the spectral bands for each pixel.}
    \label{fig:pushbroom}
\end{figure}

This study focuses specifically on anomaly detection using hyperspectral line-scan cameras, as they are an increasingly popular choice of camera for remote sensing applications. Line-scan cameras capture one line of pixels at a time and require the motion of an attached platform such as a drone, aircraft, or satellite to collect an image (Figure \ref{fig:pushbroom}). They have relatively high signal-to-noise ratios \cite{kerekes2007hyperspectral}, high spatial and spectral resolutions \cite{aasen2018linescanhighres}, and are easier and cheaper to build with commercially available components \cite{sigernes2018DIYHSI, mao2022openhsi}. Real-time line scanning is a causal time-series operation that processes each line of pixels as they are received at time $t$, with access only to previous lines \cite{chang2002anomaly, chen2014real}. Some real-time anomaly detection applications that use these cameras include search and rescue, defence surveillance and reconnaissance, and hazard detection in agricultural harvesting \cite{eismann2009automated, stellman2000real, horstrand2019novel}. 

An essential consideration for line-scan cameras is their sensitivity to unexpected platform movements and vibrations, which causes distortions that require geometric correction when creating an image from a collection of lines \cite{aasen2018linescanhighres}. Radiometric correction is also typically performed to convert raw data into physically meaningful units such as reflectance. However, geometric and radiometric corrections are complex and time-consuming, and are often unsuitable for real-time applications. Ideally, effective algorithms should function without these corrections.

Hyperspectral anomaly detection (HAD) has attracted significant research attention, with methods evolving from simpler traditional statistics and distance-based metrics to more complex unsupervised deep learning models \cite{su2021HADReview}. The HAD benchmark is the Reed-Xiaoli (RX) algorithm \cite{reed1990adaptive}, which calculates the Mahalanobis distance between each pixel and the background of a given HSI. The background is modelled as a normal distribution, represented by the mean and covariance of all pixels in this HSI. 

In recent years, many new approaches have emerged. Collaborative representation-based methods use a regularised linear model to estimate a pixel from its neighbours, identifying anomalies by high prediction errors \cite{li2014collaborative, su2018CR4HAD, wu2022CR4HAD, wang2023CR4HAD}. Sparse representation adds an additional constraint by limiting the model to estimate the pixel using a few bands, reducing noise and improving anomaly detection \cite{li2015SR4HAD, ling2018SR4HAD, lin2022SR4HAD}. Low-rank representation methods decompose the HSI into a low-rank background component, detecting anomalies that deviate from this representation \cite{xu2015LLROG, li2020LRR4HAD, wang2022LLR4HAD}. Tensor-based methods have also been used to isolate anomalies by modelling the background, but preserve the spatial and spectral structure of the image \cite{li2020Tensor4HAD, sun2022Tensor4HAD, feng2023TensorRing}. This makes them well suited to the high dimensionality of hyperspectral data. Deep learning is an area of rapid growth, with various auto-encoding techniques used to reconstruct HSIs \cite{xiang2022GuidedAE4HAD, li2023MAE4HAD, gao2023SSL4HAD, cheng2023AE4HAD, li2024DL4HAD, wang2024DL4HAD}. Autoencoders typically detect anomalies through large reconstruction errors. Many additional works combine a variety of these different methods \cite{su2020LRR+CR, wang2022LRR+DLAE, li2023LRR+DL}. Some others even use chessboard topology \cite{gao2023Chessboard}, an ``effective anomaly space" for improved anomaly-background separation \cite{chang2022EAS}, component decomposition \cite{chen2021Decomposition}, shallow graph neural networks \cite{fan2021GraphAE}, and more \cite{chang2022Sphering, chang2023IterativeSpectralSpatial}. 

However, none of the aforementioned algorithms are designed for real-time processing, let alone for the additional complexities of line scanning. Most HAD methods analyse the entire image at once, and are only demonstrated on powerful CPUs and GPUs. This requirement for high computational and memory resources makes them unsuitable for real-time applications onboard platforms with limited computing power. In general, real-time HAD is a less explored field that has an established need for more research \cite{su2021HADReview, racetin2021HADReview, xu2022HADReview, hu2022HADReview}. 

Variants of the RX algorithm have emerged as a popular choice for line-scan anomaly detection due to their effectiveness and lightweight design. They use different techniques to process the hyperspectral data pixel-by-pixel or line-by-line, making them ideal for real-time applications. One technique is the Woodbury matrix identity, which is used to recursively update the inverse correlation matrix for the Mahalanobis distance calculations \cite{chen2014real, du2009fast, liu2023hyperspectral}. Calculating the inverse is an otherwise slow and memory-intensive operation if performed for each new line. Other works apply Cholesky decomposition and forward substitution to avoid calculating the inverse covariance altogether \cite{zhao2018fast, zhao2020real}. 

Window-based methods were used to improve anomaly detection by considering only the local surroundings of each pixel \cite{kwon2003dwrx, chen2022G2LHTD, rossi2014rx, acito2013complexity, zhao2015global, liu2023RTMLMBHAD, liu2024LocalPixelwiseFMAD}, and kernel methods were found to enhance detection by better mapping the relationships between hyperspectral bands \cite{zhao2017progressive}. Algorithms using dimensionality reduction have also been proposed, using real-time versions of principal component analysis (PCA) and subspace projection \cite{horstrand2019novel, diaz2019line}. These reduce the number of spectral bands to improve the processing speed. 

However, even with these techniques, line scan anomaly detection algorithms lack a combination of three key characteristics:
\begin{itemize}
    \item Scalability - most algorithms struggle to handle high band numbers and line widths onboard small, low-power computing devices; a crucial requirement for real-time anomaly detection onboard drones or satellites.
    \item Adaptability - many algorithms do not efficiently manage changing scenery or lighting, which is essential when monitoring large and diverse locations. This is known as concept drift or non-stationary behaviour, where the underlying distribution of an image changes over time.
    \item Robustness - The literature has overlooked demonstrating algorithm stability and performance on larger datasets, particularly with the geometric distortions of line scanning. Previous studies often assessed algorithms on small, geometrically corrected datasets such as the Airport-Beach-Urban (ABU) dataset \cite{kang2017ABUDataset} or San Diego Airport, with only a few hundred pixels in height and width. Algorithms need testing on complex datasets to assess their suitability for real-time remote sensing applications.
\end{itemize}

To overcome these challenges, the Exponentially moving RX algorithm (ERX) is proposed. ERX processes hyperspectral images line-by-line, simultaneously updating the underlying background model and detecting anomalies. Sparse random projection (SRP) is used to reduce the dimensionality of the raw hyperspectral bands, allowing ERX to scale to high dimensions. ERX also uses exponentially moving averages (EMA) to rapidly update the mean and covariance, and quickly adapt to changing scenery by prioritising recent information. Cholesky decomposition is also used to further improve performance. ERX demonstrates robustness by outperforming similar algorithms on larger, more complex datasets that include geometric distortion and changing scenery. In summary, the overall contributions of this work are:
\begin{enumerate}
    \item A thorough review of RX-based anomaly detection algorithms for line scanning. These algorithms are grouped by similar methods to provide an intuitive summary of the related work.
    \item The Exponentially moving RX anomaly detector (ERX). ERX employs sparse random projection, exponential moving averages, and Cholesky decomposition to make it fast, scalable, adaptable, and robust. ERX outperforms similar algorithms in both speed and detection when evaluated on a small edge computer.
    \item Three novel datasets for anomaly detection. These datasets are substantially larger than those used in the literature and include examples of changing scenery and geometric distortions. These datasets better capture the practical challenges of real-time anomaly detection for hyperspectral line scanning.
    \item HyperAD: \href{https://github.com/WiseGamgee/HyperAD}{https://github.com/WiseGamgee/HyperAD}, an openly available Python repository that contains the implementation of ERX and other algorithms from the literature. This is the first open source repository for real-time anomaly detection for hyperspectral line scanning, which promotes accessibility and future work. 
\end{enumerate}

The remainder of this paper is structured as follows. Section \ref{sec:related_work} reviews related work on real-time anomaly detection for hyperspectral line scanning. Section \ref{sec:erx} details the proposed method, ERX. Section \ref{sec:experiments} covers the datasets and experimental design. Section \ref{sec:discussion} presents and discusses the results, outlining the core findings, limitations, and future directions. Finally, Section \ref{sec:conclusion} concludes the paper.

\section{Related Work} \label{sec:related_work}

This section reviews RX-based algorithms that are used for real-time anomaly detection in line scanning. These algorithms are online, meaning that they detect anomalies while updating their models. They are separated into subsections that share common approaches.

\subsection{The RX Algorithm}

The RX algorithm \cite{reed1990adaptive} detects anomalous pixels with spectral signatures that noticeably differ from the background of a given hyperspectral image, which is modelled as a normal distribution. The background is defined by the mean pixel vector \eqref{eqn_rx_mean} and covariance matrix \eqref{eqn_rx_cov}, calculated using all pixels in the image:
\begin{align}
    \bm{\mu} &= \frac{1}{n} \sum^{n}_{i = 1} \bm{x}_i \label{eqn_rx_mean} \\
    \bm{K} &= \frac{1}{n} \sum^{n}_{i = 1} (\bm{x}_i - \bm{\mu}) (\bm{x}_i - \bm{\mu})^T \label{eqn_rx_cov}
\end{align}
\noindent
where $ \bm{x}_i \in \mathbb{R}^{b} $ represents each pixel, $b$ is the number of hyperspectral bands, and $n$ is the total number of pixels in the image. The Mahalanobis distance \eqref{eqn_rx_md} is then calculated for each pixel:
\begin{align} 
    \delta_{i} &= \sqrt{(\bm{x}_i - \bm{\mu}) \bm{K}^{-1} (\bm{x}_i - \bm{\mu})^T} \label{eqn_rx_md}
\end{align}

The Mahalanobis distance gives an estimate of how far each pixel lies from the background, where a larger distance means a greater likelihood of being an anomaly. Finally, pixels are classified as anomalies if their distance is above a user-defined threshold ($\tau$):
\begin{align}
    \hat{y}_i &= \begin{cases}
                    1, & \text{if } \delta_{i} \geq \tau \\
                    0, & \text{otherwise}
                  \end{cases} \label{eqn_rx_threshold}
\end{align}

A core limitation of the base RX algorithm is that it requires the capture and storage of every single pixel before calculating the mean and covariance. This prevents real-time detection and easily results in insufficient memory due to the large amount of data. Calculating the inverse of the covariance matrix is also a slow operation because of the large number of hyperspectral bands. A common solution to these issues is to perform recursive mean and covariance updates, of which one popular method is the Woodbury Matrix Identity.

\subsection{The Woodbury Matrix Identity}

The Woodbury matrix identity \eqref{woodbury_vector} is an expression that has been used to recursively update the inverse covariance matrix as new lines are captured from the camera, reducing memory usage and processing time.
\begin{align} 
[\bm{A} + \bm{u}\bm{c}\bm{v}^T]^{-1} = \bm{A}^{-1} - \frac{[\bm{A}^{-1} \bm{u}][\bm{v} \bm{A}^{-1}]}{\bm{c}^{-1} + \bm{v} \bm{A}^{-1} \bm{u}} \label{woodbury_vector} 
\end{align}

Chen et al. \cite{chen2014real} presented the real-time causal RXD algorithm, RT-CK-RXD. RT-CK-RXD detects anomalies pixel-wise, which means that it iterates over each pixel in a new line. Chen et al. redefined the current covariance estimate $\bm{K}_t$ \eqref{eqn_rx_cov} as a function of its previous value $\bm{K}_{t-1}$ and the current pixel $\bm{x}_t$, shown in equation \eqref{eqn_chen_cov1}:
\begin{align} 
 \bm{K}_t &= \frac{n - 1}{n}\bm{K}_{t-1} + \frac{1}{n} (\bm{x}_t - \bm{\mu}_t) (\bm{x}_t - \bm{\mu}_t)^T \label{eqn_chen_cov1}
\end{align}
\noindent
where $n$ is the total number of pixels processed at time $t$. By inverting \eqref{eqn_chen_cov1}, the inverse covariance was then equated to the LHS of the Woodbury Matrix identity equation in \eqref{woodbury_vector}, giving:
\begin{align} 
 \bm{K}_t^{-1} &= [\frac{n - 1}{n}\bm{K}_{t-1} + \frac{1}{n} (\bm{x}_t - \bm{\mu}_t) (\bm{x}_t - \bm{\mu}_t)^T]^{-1} \label{chen_cov2}
\end{align}

Given that $\bm{A} = \frac{n - 1}{n}\bm{K}_{t-1}$, $\bm{u} = \bm{v} = \frac{1}{\sqrt{n}} (\bm{x}_t - \bm{\mu}_t)$, and $c=1$, these values are substituted into the RHS of \eqref{woodbury_vector} and the inverse covariance matrix is updated.  

Chen et al. \cite{chen2014real} also define RT-CR-RXD, which replaces the covariance matrix with the correlation matrix $ \bm{R}_t = \frac{1}{n} \sum^{n}_{i = 1} \bm{x}_i \bm{x}_i^T$. This algorithm is a faster alternative, as the correlation matrix does not require demeaning of each pixel while still extracting the correlation between the hyperspectral bands.

Du and Nekovei \cite{du2009fast} proposed several in-depth RX algorithms for real-time anomaly detection. One algorithm, RX-BIL, used the generalised Woodbury matrix identity to update the inverse correlation matrix for each new line $\bm{X_t}$:
\begin{align}
    \bm{R}^{-1}_t &= \bm{R}^{-1}_{t-1} - \frac{[\bm{R}^{-1}_{t-1} \bm{X}_t][\bm{X}_t^T \bm{R}^{-1}_{t-1}]}{\bm{I} + \bm{X}_t^T \bm{R}^{-1}_{t-1} \bm{X}_t} \label{BIL_eq}
\end{align}

RX-BIL randomly removed pixels to increase detection speed. More pixels can be removed in homogeneous images as fewer are needed to accurately estimate the background mean and covariance. Fewer pixels should be removed in diverse images, so that more variable spectra are captured by the background statistics.

Although the Woodbury matrix identity reduces computation time, each line equally contributes to the background estimation regardless of its collection time. Therefore, it does not inherently adapt to changing scenery and risks detecting false anomalies from redundant data. Liu et al. \cite{liu2023hyperspectral} partially addressed this issue using a combined Woobury matrix identity and finite Markov model for the line-by-line anomaly detection algorithm FMLRT-RAD. FMLRT-RAD improved stability and adaptability by using only recent lines, demonstrating slightly higher detection performance than RT-CR-RXD. However, its processing speed remained comparable, indicating a need for further speed improvements.

\subsection{Cholesky Decomposition}

Cholesky decomposition paired with forward substitution avoids calculating the inverse covariance altogether, offering a faster alternative to the Woodbury matrix identity. Zhang et al. \cite{zhang2017fast} proposed CDLSS, a line-by-line anomaly detection algorithm that used Cholesky decomposition. Given a new line of pixels $\bm{X}_t$, the corresponding Mahalanobis distance of each pixel defined as:
\begin{align}
    \delta_i &= \sqrt{\bm{x}_i \bm{R}_{t-1}^{-1} \bm{x}_i^T \label{cdlss_dm_start}}
\end{align}
\noindent
where $\bm{x}_i$ is the $i$th pixel vector in line $\bm{X}_t$, and $\bm{R}_{t-1}$ is the correlation matrix in the previous time step. The correlation matrix is decomposed into its lower triangle matrix representation $\bm{R}_{t-1} = \bm{L}_{t-1} \bm{L}_{t-1}^T$, which can be substituted into \eqref{cdlss_dm_start}:
\begin{align}
    \delta_i &= \sqrt{(\bm{x}_t \bm{L}_{t-1}^{-1}) (\bm{x}_t \bm{L}_{t-1}^{-1})^T}
\end{align}

By substituting in $(\bm{X}_t \bm{L}_{t-1}^{-1}) = \bm{m}_i$, the Mahalanobis distance is represented as:
\begin{align}
    \delta_i &= \sqrt{\bm{m}_i \cdot \bm{m}_i} \label{cdlss_dm_end}
\end{align}

As $\bm{x}_i$ is known and $\bm{L}_{t-1}$ is calculated from Cholesky decomposition, forward substitution is used to calculate $\bm{m}_i$ directly, given that $\bm{m}_i \bm{L}_{t-1} = \bm{x}_i$. However, using the correlation matrix increases the instability of the algorithm because of the higher risk of singularities, which can cause it to fail because it cannot compute the inverse.

Zhao and Xi-Feng \cite{zhao2018fast} presented LRT-KRXD-CD, using Cholesky decomposition to reduce the processing time of the local real-time kernel RX detector (LRT-KRXD) \cite{zhao2016real}. Zhao et al. \cite{zhao2020real} presented RT-KCRD, building on KCRD \cite{li2014collaborative} with Cholesky decomposition. RT-KCRD used a moving window (discussed in the following section), recursive updates for the kernel covariance matrix, and an optimised regularisation matrix using previous anomaly scores. RT-KCRD was on average 86 times faster than KRXD and KCRD, and was comparable to LRT-KRXD-CD \cite{zhao2020real}. Although these algorithms demonstrated the speed-up potential of Cholesky decomposition, they remain relatively slow due to the kernel function (discussed in Section \ref{lr: kernel-methods}). Overall, Cholesky decomposition has yet to be used in a robust and fast algorithm.

\subsection{Local/Window Algorithms}

\begin{figure}[h!]
    \centering
    \includegraphics[width=\linewidth]{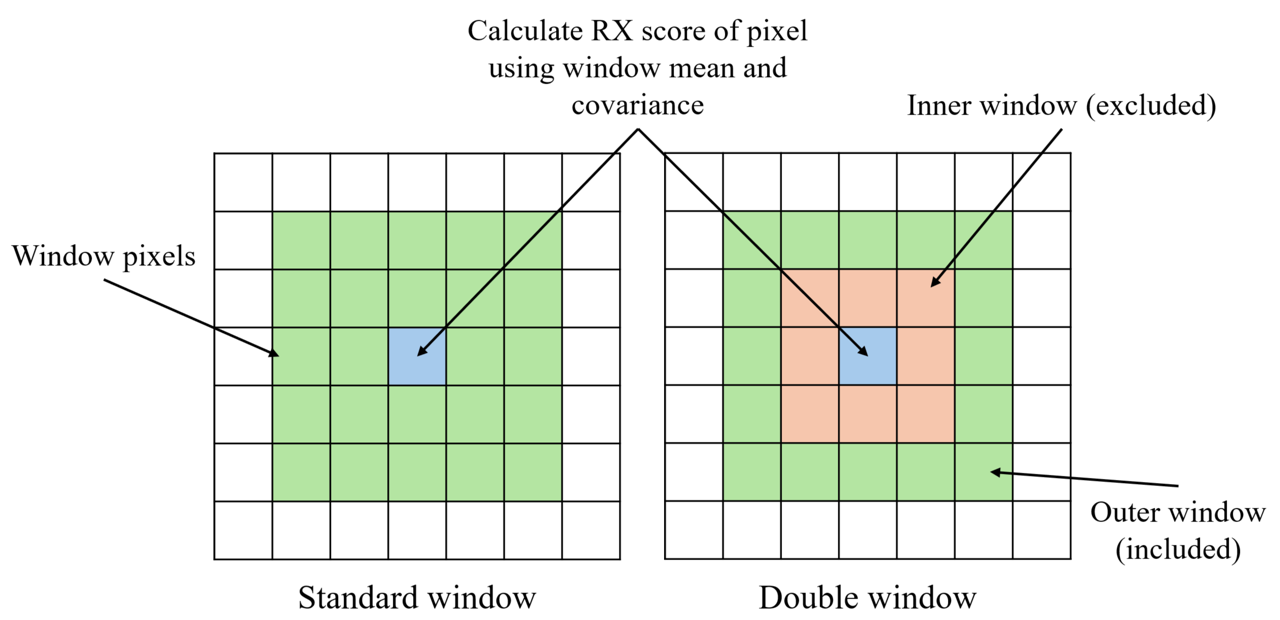}
    \caption{A standard window (left) versus a double window (right). The central pixel (blue) is analysed, with each approach estimating the local mean and covariance from the surrounding pixels.}
    \label{fig:window}
\end{figure}

Window-based algorithms extract anomalies locally rather than at a global level, improving detection performance \cite{kwon2003dwrx, chen2022G2LHTD} and adaptability. As shown in Figure \ref{fig:window}, a standard window is a small region surrounding the current pixel being processed. Double-windows discard a small inner region of pixels, minimising the contribution of the same object to the local background statistics. 

Rossi et al. \cite{rossi2014rx} improved two local anomaly detection algorithms using fast updates of the inverse covariance. The first, RX-LBL-IU, builds upon the work of Acito et al. \cite{acito2013complexity}, which used ``sliding" double-windows around each pixel to identify anomalies. The sliding window approach used the Woodbury matrix identity to recursively update the window inverse covariance, instead of recalculating it for each pixel's window. The second method is Dark HORSE RX \cite{stellman2000real}, which used an exponentially rolling mean and covariance for pixel-wise updates. Rossi et al. \cite{rossi2014rx} deemed RX-LBL-IU the most effective of the two algorithms.

Zhao et al. \cite{zhao2015global} presented a causal pixel-wise approach, only accessing the window pixels captured before the current pixel being analysed. A two-step Woodbury matrix identity was used to update the inverse correlation matrix by adding the latest pixel and removing the oldest one. Liu et al. introduced two more local anomaly detection algorithms; one designed around multi-line multi-band processing \cite{liu2023RTMLMBHAD} and another for pixel-by-pixel processing \cite{liu2024LocalPixelwiseFMAD}. These algorithms update the correlation matrix recursively and focus on recent pixels.

However, calculating the window statistics efficiently is a major challenge. Small window sizes can lead to matrix singularities, and large window sizes lead to slow compute times. Preliminary experiments with Python were too slow for real-time detection, failing to meet the speed requirements. Many window approaches also assume geometrically corrected pixels, which can reduce the robustness for real-time line scanning.

\subsection{Kernel Methods} \label{lr: kernel-methods}

Kernel methods address the assumption of a normal distribution in the RX algorithm and subsequently improve detection accuracy. A kernel function allows the RX algorithm to map the non-linear correlations that hyperspectral bands share with each other in practice \cite{kwon2005kernel}. PLP-KRXD was proposed by Zhao et al. \cite{zhao2017progressive}, using parallel sliding windows and the Woodbury matrix identity to speed up the KRX algorithm \cite{kwon2005kernel}. The sliding windows were implemented as a faster alternative to the double-window model. The Woodbury matrix identity reduced the time required to calculate the inverse covariance of multiple sliding windows. Zhao et al. demonstrated that PLP-KRXD achieved detection performance similar to KRXD, but was 40 to 58 times faster, depending on the dataset. 

Kernel methods, though more accurate, are much slower due to the additional matrix operations required. This work focuses on non-kernel methods which are better suited to real-time processing with limited compute resources.

\subsection{Dimensionality Reduction}

Dimensionality reduction methods improve processing speed and signal-to-noise ratio by transforming hyperspectral bands into fewer features. Horstrand et al. \cite{horstrand2019novel} developed LbL-AD, which uses Principal Component Analysis (PCA) to reduce the number of bands. Line-by-line subspace calculation using the power method and deflation was used to efficiently approximate the eigenvalues $\bm{e}_t$ and the eigenvectors $\bm{E}_t$. Using orthogonal subspace projection, the dimensions of the hyperspectral line were reduced:
\begin{align}
    \bm{Z}_t &= \bm{E}_t^T \bm{X}_t \label{subspace_proj_eqn}
\end{align}
\noindent
where $\bm{X}_t$ is the most recent hyperspectral line and $\bm{Z}_t$ is the line after the dimensionality reduction, with the eigenvalues ${e}_t$ forming the diagonal of the lower-dimensional covariance matrix. The Mahalanobis distance was calculated as:
\begin{align}
    \delta_i &= \sqrt{\bm{z}_i \text{diag}(\bm{e}_t)^{-1} \bm{z}_i^T} \label{pca_md_eqn}
\end{align}
\noindent
where $\bm{z}_i$ is the $i$th dimensionality-reduced pixel in the current line. The authors introduced an adaptive approach that prevents anomalous pixels from contributing to the background statistics. The authors recommend reducing the bands to a maximum of five principal components, as any more significantly slows down processing.

LbL-FAD was also proposed, using a modified Gram-Schmidt method for orthogonal subspace projection \cite{diaz2019line}. This algorithm has also been adapted for hardware through field-programmable gate arrays (FPGAs), demonstrating the efficiency of the algorithm onboard small platforms in real time \cite{diaz2021fpga, caba2022low}. These dimensionality reduction algorithms are fast, accurate, and scalable. However, like the vast majority of related work, they are typically evaluated on small, uniform, and geometrically corrected HSI datasets. These datasets do not reflect the changing scenery and distortions encountered by drones or satellites when surveying large areas in real time.

\subsection{Summary}

Existing methods employ techniques such as the Woodbury matrix identity, Cholesky decomposition, local windows, kernel functions, and dimensionality reduction to enhance the RX algorithm for line scanning. However, they were found to lack one or more key requirements: scalability for processing large numbers of HSI bands and pixels in real time, adaptability to changing scenery, and robustness to geospatial distortions. Previous work also focused on small, corrected image datasets with limited scenery variation, revealing an additional gap in evaluating line-scanning algorithms under practical conditions. In the following section, we present ERX and three larger and more diverse datasets to address these challenges. Two datasets have changing scenery to test adaptability, with one also having geometric distortions to better evaluate robustness. 

\section{Proposed Algorithm - ERX} \label{sec:erx}

\subsection{Overview}

\begin{figure*}
    \centering
    \begin{minipage}{\linewidth}
        \centering
        \includegraphics[width=\linewidth]{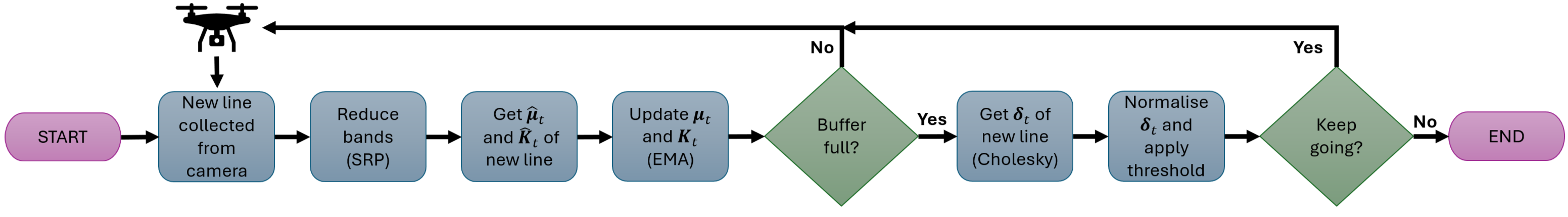}
    \end{minipage}
    \caption{ERX algorithm flowchart. SRP is sparse random projection, and $\bm{\hat{\mu}}_t$ and $\bm{\hat{K}}_t$ are the mean and covariance of the current line. $\bm{\mu}_t$ and $\bm{K}_t$ are the background mean and covariance, and EMA refers to the exponentially moving averages. $\bm{\delta}_t$ is the Mahalanobis distance vector of all pixels in the line.}
    \label{fig:erx_flowchart}
\end{figure*}

The Exponentially moving RX algorithm (ERX) is an online algorithm, which means that it detects anomalies while simultaneously updating itself in real time. Figure \ref{fig:erx_flowchart} illustrates the sequential processing of hyperspectral lines by ERX, one line at a time. ERX features two main components: sparse random projection (SRP) and exponentially moving averages (EMAs). SRP enhances scalability by reducing the number of hyperspectral bands to a smaller number of dimensions. This significantly enhances the processing speed, as the number of bands is a major performance bottleneck for real-time deployment. Line-by-line EMAs rapidly update the background mean and covariance, prioritising recent lines and adapting to changing scenery. 

Initially, a small buffer period is used to improve the modelling of the background mean and covariance before starting detection. For this research, the buffer is set to 99 lines, but in practice ERX can start detection immediately without storing previous hyperspectral lines. Cholesky decomposition and forward substitution are used to calculate the Mahalanobis distance of each pixel, with minor stability improvements over conventional methods. The algorithm concludes with a line-by-line normalisation of the Mahalanobis distances to identify anomalous pixels. This process is repeated for any number of hyperspectral lines. ERX's Python implementation is openly available under ``detectors" in the GitHub repository. The remainder of this section expands on the main components of ERX.

\subsection{Sparse Random Projection (SRP)}

The large number of hyperspectral bands is one of the biggest drags on performance and scalability for real-time anomaly detection. Increasing the number of spectral bands disproportionately decreases the processing speed and exposes algorithms to the Hughes phenomenon \cite{signoroni2019CursedHSI}, also known as the curse of dimensionality. This means that the pixels are more spread out in the feature space, making the separation of anomalies from the background more difficult. To alleviate these issues, ERX uses sparse random projection (SRP) \cite{achlioptas2003SparseRandomProjections, li2006VerySparseRandomProjections} to reduce the original hyperspectral bands to a smaller number of dimensions. 

Random projection reduces the input dimensionality using a matrix of randomly generated weights \cite{bingham2001RandomProjection}. SRP zeroes many of these weights to further reduce the computational load. SRP is simpler and more efficient compared to other methods such as PCA \cite{dasgupta2000RandomProjection}, and still preserves the relative distances between the original data. SRP has been used to reduce the dimensionality of text and image data \cite{bingham2001RandomProjection}, but this work appears to be the first use in hyperspectral anomaly detection.

In the context of ERX, the random weight matrix $\bm{W} \in \mathbb{R}^{b,d}$ is generated during initialisation with the following distribution ($b$ is the number of bands and $d$ is the new number of dimensions after projection):
\begin{align}
w_{i,j} &= \frac{\sqrt{s}}{\sqrt{d}} \begin{cases}
                                         1, & \text{with probability } \frac{1}{2s} \\
                                         0, & \text{with probability } 1 - \frac{1}{s} \\
                                         -1, & \text{with probability } \frac{1}{2s} 
                                     \end{cases} \label{erx_threshold}
\end{align}
\noindent
where $w_{i,j}$ is any weight in $\bm{W}$, and the sparsity parameter $s=\sqrt{b}$ \cite{li2006VerySparseRandomProjections}. The number of active (non-zero) weights in the random projection matrix equals the number of matrix elements multiplied by the activation probability:
\begin{align}
    n_{aw} &= (b \times d) \times \frac{1}{\sqrt{b}} \\
    &= d\sqrt{b}
\end{align}

Given the latest hyperspectral line $\bm{X}_{t} \in \mathbb{R}^{p,b}$ (where $b$ is the number of bands, $p$ is the number of pixels per line), every pixel is projected to the lower number of dimensions ($d$):
\begin{align}
    \bm{Z}_{t} = \bm{X}_{t} \cdot \bm{W} 
\end{align}
\noindent
where $\bm{Z}_{t} \in \mathbb{R}^{p,d}$ is the reduced hyperspectral line with $d$ dimensions. 

The Johnson-Lindenstrauss Lemma is typically used to determine the smallest number of dimensions \cite{achlioptas2003SparseRandomProjections} that preserves the original distances between the spectral signatures of the original HSI. Unfortunately, this approach requires an initial image sample to estimate the optimal $d$, which is not available for this task. Hence, $d$ is empirically set as $5$ for all tests, with the effect of altering $d$ explored in Section \ref{subsec:ablation_study}. 

Empirical dimension selection also improves flexibility by removing the dependency on an initial background sample, enabling adaptation to changing backgrounds. This contrasts with the existing algorithms that use dimensionality reduction methods, which assume a uniform or constant background. This implementation of SRP improves processing speed, reduces memory usage, and boosts the signal-to-noise ratio without any optimisation. However, random feature selection introduces some variability in model performance, which is investigated through repeated experiments. The SRP implementation of the scikit-learn library is used \cite{sklearn}.

\subsection{Exponentially Moving Mean \& Covariance}

Existing algorithms, such as those that use the Woodbury matrix identity, update the background statistics incrementally or through methods such as simple moving averages (SMAs). Incremental updates can be fast, but treat all pixels equally which limits adaptability to scene changes. SMAs provide adaptability by focusing on a smaller window of pixels, though they operate more slowly. ERX uses exponentially moving averages (EMAs) to rapidly update the background mean and covariance estimates, line-by-line. This fast operation prioritises recent lines, directly enhancing adaptability. 

\begin{figure}
    \centering
    \includegraphics[width=\linewidth]{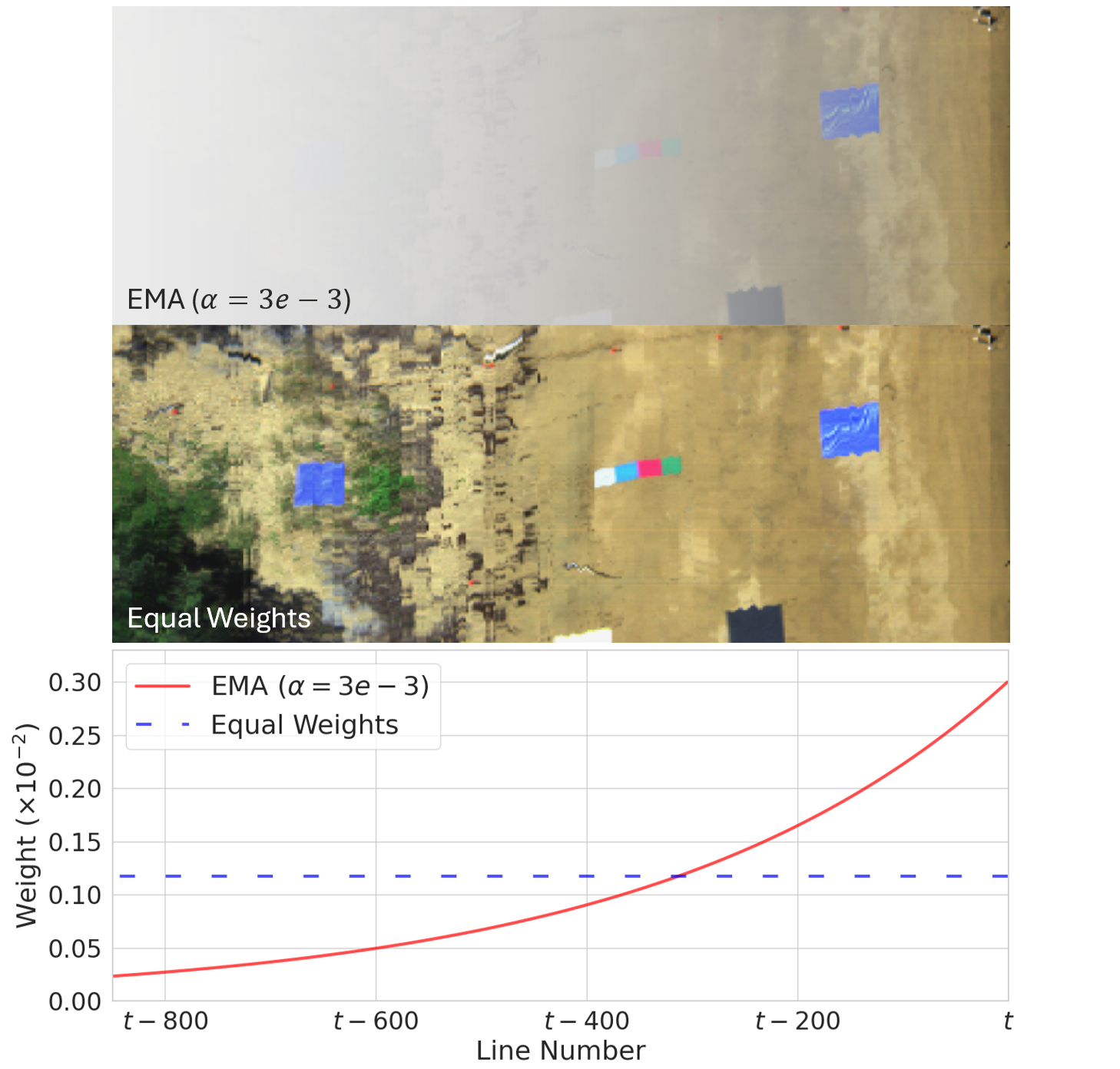}
    \caption{Conceptual comparison of ERX's exponentially moving averages (EMAs), versus equal weighting. EMAs prioritise newer lines and forget older ones (top), unlike equally weighted methods, which treat all lines uniformly (middle). The plot (bottom) displays the weight profiles for all lines in the example scene, with the most recent captured at time $t$.}
    \label{fig:exponential_vision}
\end{figure}

This concept is demonstrated in Figure \ref{fig:exponential_vision}, where the use of an exponential momentum factor ($\alpha$) is used to change the ``memory" of the algorithm. Increasing the momentum parameter $\alpha$ increases the weighting of recent lines, shortening ERX's memory and making older lines less relevant for anomaly detection. This allows ERX to adapt to changing scenery in real time. This seems to be the first use of line-by-line exponentially moving statistics for HAD.

To implement this, the mean vector ($\bm{\hat{\mu}}_{t} \in \mathbb{R}^{d}$) and covariance matrix ($\bm{\hat{K}}_{t} \in \mathbb{R}^{d,d}$) of the projected hyperspectral line ($\bm{Z}_{t}$) are calculated: 
\begin{align}
    \bm{\hat{\mu}}_{t} &= \frac{1}{p} \sum\limits_{i=1}^{p} \bm{z}_i \\
    \bm{\hat{K}}_{t} &= \frac{1}{p-1} \sum\limits_{i=1}^{p} (\bm{z}_i - \bm{\hat{\mu}}_{t})(\bm{z}_i - \bm{\hat{\mu}}_{t})^T
\end{align}
\noindent
where each pixel in $\bm{Z}_{t}$ is defined as $\bm{z}_i \in \mathbb{R}^d$ ($d$ is the number of dimensions after SRP, $p$ is the number of pixels per line, and $ i = 1, 2, ..., p $). The background mean ($\bm{\mu}_t$) and covariance ($\bm{K}_t$) are then updated using the momentum factor $ \alpha $ (where $ 0 < \alpha < 1 $):
\begin{align}
    \bm{\mu}_t &= (1 - \alpha) \bm{\mu}_{t-1} + \alpha \bm{\hat{\mu}}_{t} \\
    \bm{K}_t &= (1 - \alpha) \bm{K}_{t-1} + \alpha \bm{\hat{K}}_{t}
\end{align}

The default momentum value is empirically set to $\alpha=0.1$ for most experiments, and the initial background values when $t=0$ are given by the mean and covariance of the first line. Momentum can affect ERX's detection performance and is examined in detail in Section \ref{subsec:ablation_study}.

\subsection{Cholesky Decomposition \& Improved Stability}

Like Zhang et al. \cite{zhang2017fast}, Cholesky decomposition is used to quickly compute the Mahalanobis distance by avoiding matrix inversion. Due to the correlation matrix's instability and singularity risk, ERX simply uses the covariance matrix $\bm{K}_t^{-1}$ plus a small identity matrix $10^{-5}\bm{I}$ instead:
\begin{align}
       \delta_{t} &= \sqrt{[(\bm{z}_{i}-\bm{\mu}_t) \bm{L}_t^{-1}] [(\bm{z}_{i}-\bm{\mu}_t) \bm{L}_t^{-1}]^T} \label{erx_choleskied}
\end{align}
\noindent
where $\bm{L}_t$ is the lower triangle of $\bm{K}_t^{-1} + 10^{-5}\bm{I}$. Adding a small identity matrix enhances the stability of the algorithm by ensuring a positive definite covariance matrix, resolving failures found in previous algorithms. \eqref{erx_choleskied} can be simplified by substituting $\bm{m}_i=(\bm{z}_{i}-\bm{\mu}_t) \bm{L}_t^{-1}$:
\begin{align}
    \delta_{i} &= \sqrt{\bm{m}_i \cdot \bm{m}_i^T} \label{erx_end}
\end{align}
\noindent
where $\bm{m}_i$ is calculated directly using forward substitution for $\bm{L}_t\bm{m}_i =(\bm{z}_{i}-\bm{\mu}_t)$. This approach provides a more robust solution when using Cholesky decomposition.

\subsection{Line-by-line Normalisation}

Using a threshold is the final step in anomaly detection, distinguishing Mahalanobis distances above the boundary as anomalies and those below as normal. ERX first normalises the Mahalanobis distance vector for better anomaly-background separation:
\begin{align}
    \delta_{N,i} &= \frac{\delta_i - \bar{\delta}}{\sigma_{\delta}} \label{eqn_erx_norm} 
\end{align}
\noindent
where $\bar{\delta}$ is the mean Mahalanobis distance and $\sigma_{\delta}$ is the standard deviation. Pixels are labelled anomalies if they exceed the threshold $\tau_N$: 
\begin{align}
\hat{y}_i &= \begin{cases}
                    1, & \text{if } \delta_{N,i} \geq \tau_N \\
                    0, & \text{otherwise}
                  \end{cases} \label{erx_threshold}
\end{align}

\section{Experimental Setup} \label{sec:experiments}

This section includes the datasets, experimental design, and the results. The datasets and the Python code for all algorithms and their experiments are available at \href{https://github.com/WiseGamgee/HyperAD}{https://github.com/WiseGamgee/HyperAD}.

\subsection{Datasets} \label{section_datasets}

\begin{figure*}[h!]
    \centering
    \begin{minipage}{\linewidth}
        \centering
        \includegraphics[width=\linewidth]{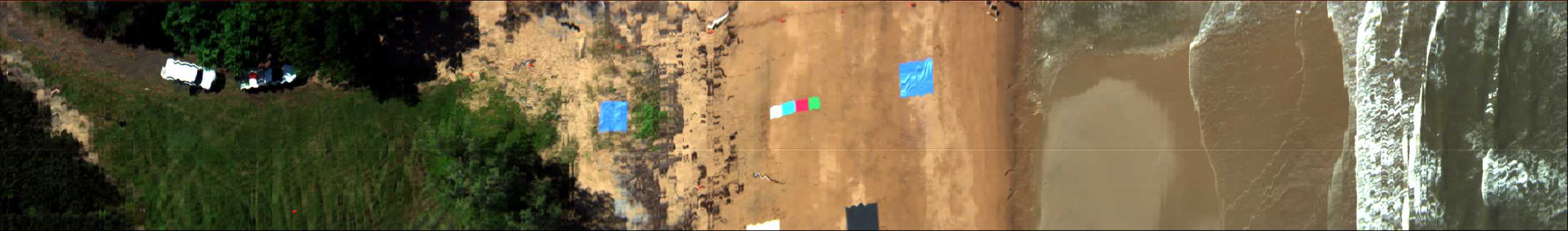}
    \end{minipage}
    %\vspace{-\baselineskip} % Remove space between figures
    \begin{minipage}{\linewidth}
        \centering
        \includegraphics[width=\linewidth]{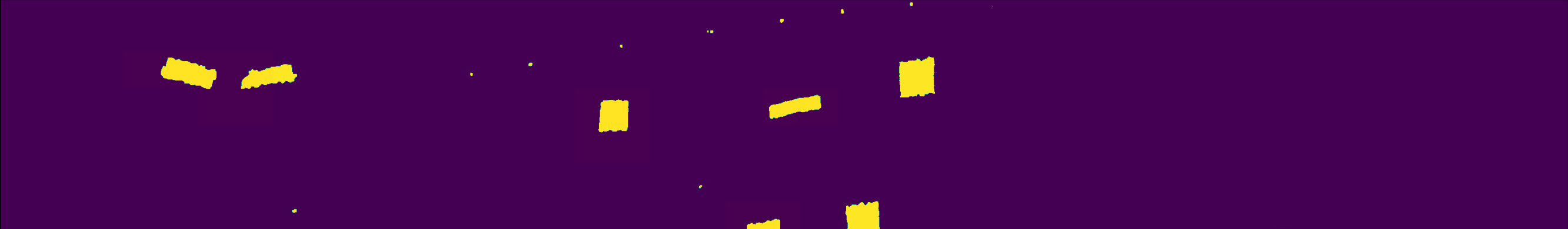}
    \end{minipage}
    \caption{The beach dataset RGB image (top) and anomaly ground truth image (bottom).}
    \label{dataset_beach}
\end{figure*}

\begin{figure}[h!]
    \centering
    \begin{minipage}{\linewidth}
        \centering
        \includegraphics[width=\linewidth]{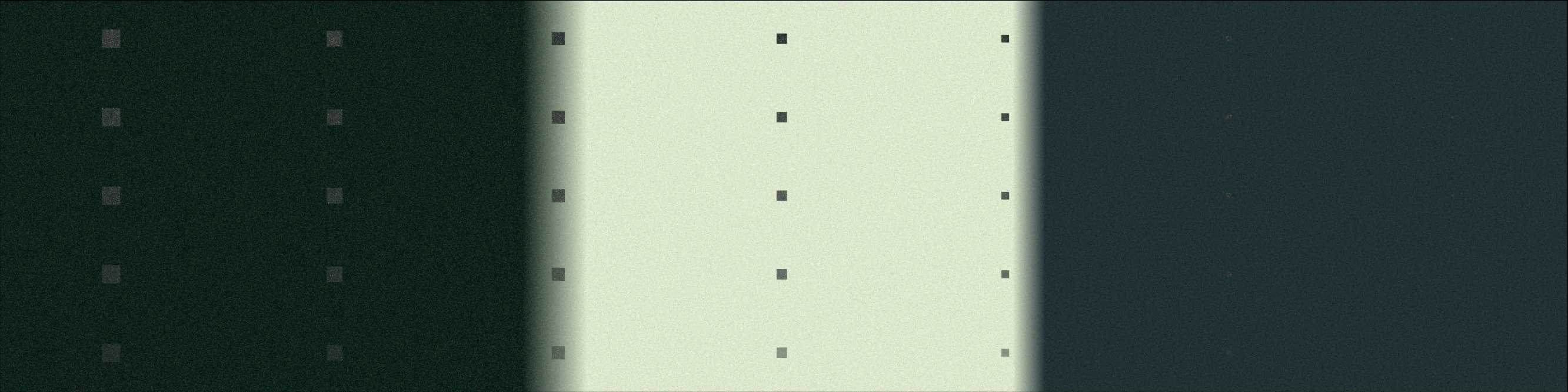}
    \end{minipage}
    %\vspace{-\baselineskip} % Remove space between figures
    \begin{minipage}{\linewidth}
        \centering
        \includegraphics[width=\linewidth]{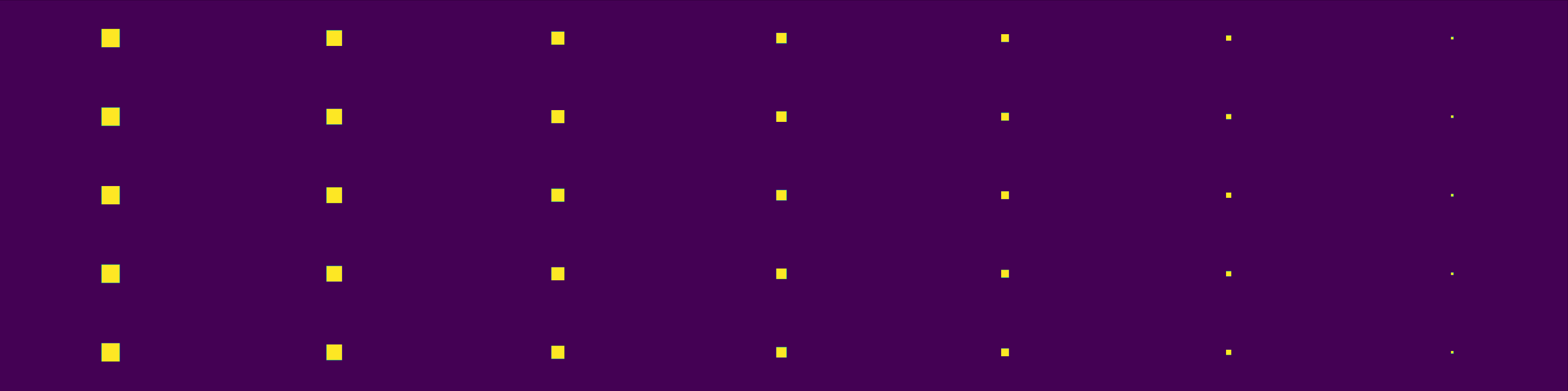}
    \end{minipage}
    \caption{The synthetic dataset RGB image (top) and anomaly ground truth image (bottom).}
    \label{dataset_synthetic}
\end{figure}

\begin{figure}[h!]
    \centering
    \begin{minipage}{\linewidth}
        \centering
        \includegraphics[width=\linewidth]{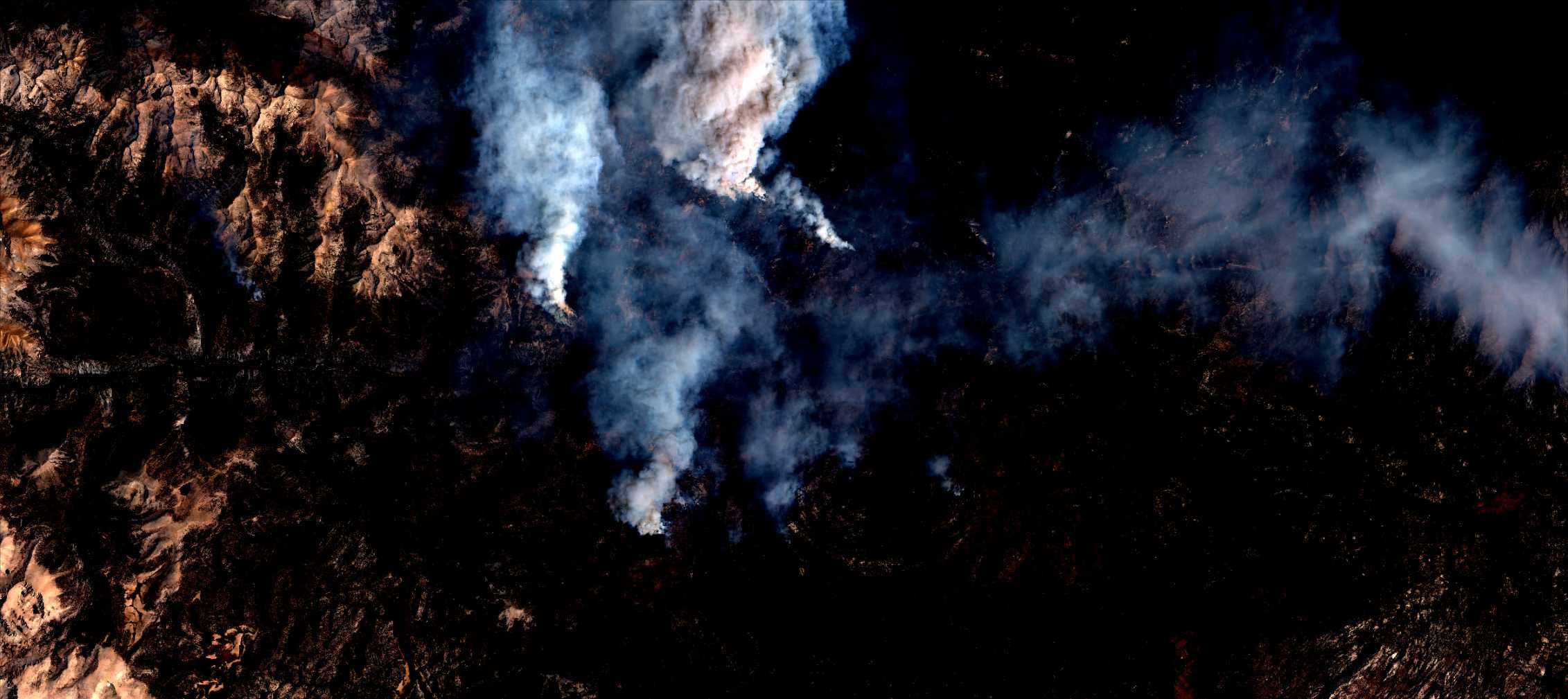}
    \end{minipage}
    %\vspace{-\baselineskip} % Remove space between figures
    \begin{minipage}{\linewidth}
        \centering
        \includegraphics[width=\linewidth]{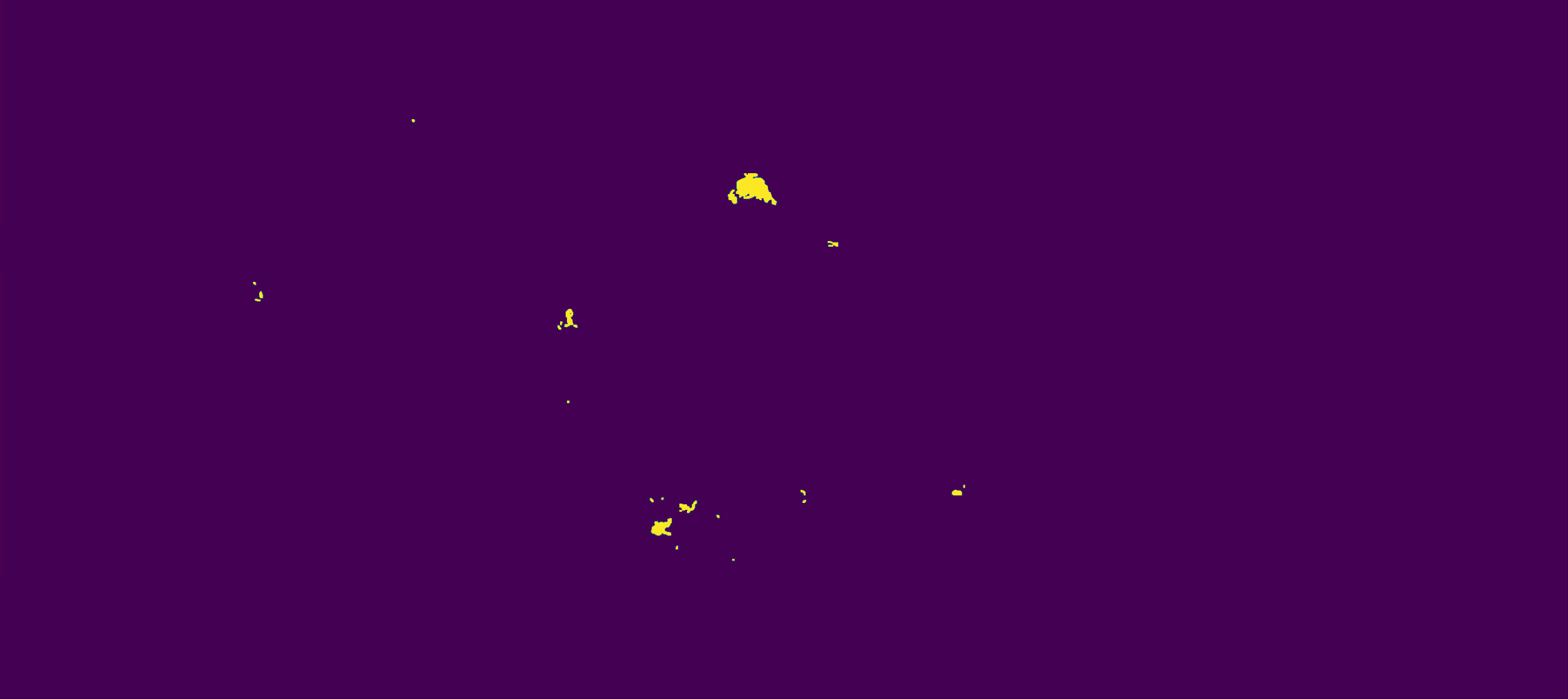}
    \end{minipage}
    \caption{The Sequoia National Park (SNP) dataset RGB image (top) and anomaly ground truth image (bottom). The dataset has few anomalous pixels, making it highly imbalanced.}
    \label{dataset_snp}
\end{figure}

Three datasets are used to evaluate the performance of each algorithm, being fed line-by-line to simulate the capture of a line-scan camera (from left to right for each image):
\begin{itemize}
    \item Beach dataset - a natural coastal area with a variety of human-made anomalies present (Figure \ref{dataset_beach}).
    \item Synthetic dataset - a constructed dataset using pixels sampled from AVIRIS with anomalous targets (Figure \ref{dataset_synthetic}).
    \item Sequoia National Park (SNP) dataset - a national park with live wildfires present (Figure \ref{dataset_snp}).
\end{itemize}

The beach dataset, sized 452 pixels by 3,072 lines with 108 bands, was collected via a drone-mounted hyperspectral line-scan system on a beach in Queensland, Australia. The resolution of the pixels is 5 cm. It transitions from vegetation, to sand, to shallow ocean, and includes anomalous objects such as different coloured and sized tarps, two vehicles, and reflective cones (the smaller yellow objects in the ground truth). This version uses radiance and lacks geometric correction, as is evident from the ``wobble" between lines in Figure \ref{dataset_beach}. For detailed dataset collection methods, see Mao et al. \cite{mao2022openhsi}.

The synthetic dataset, sized 600 pixels by 2,400 lines with 90 bands, was created by sampling 7.1m pixels from an AVIRIS radiance dataset collected near Gulfport, Mississippi \cite{green1998aviris}. The background is generated from vegetation, sand, and water pixels with smoothed transitions between each type. The anomalies are square targets made from airport runway pixels. The targets are equally spaced with decreasing size along-track (left to right), and are repeated from top to bottom with 10\%, 20\%, 30\%, and 50\% background mixing to add noise. The original 224 bands are reduced to 189 by removing the water absorption and low signal-to-noise ratio bands (as in Głomb and Romaszewski \cite{glomb2020anomaly}). After this, only the first 90 bands are kept to further improve the signal-to-noise ratio and avoid the OpenBLAS threading issue (discussed in Section \ref{sec:speed_comp}).

The Sequoia National Park (SNP) dataset, 1,116 pixels by 2,499 lines with 13 bands, was collected using the Sentinel-2 multispectral satellites and downloaded via Sentinel Hub. It has a 30m pixel resolution. The dataset covers part of Sequoia National Park in California and includes live wildfires as anomalies that were identified using infrared bands. It uses L1C processing for top-of-atmosphere reflectance, unlike the hyperspectral datasets that use radiance. This multispectral dataset helps to demonstrate algorithm performance on less correlated lower-dimensional data.

\subsection{Anomaly Detection Algorithms}

The following algorithms are compared in the experiments: 
\begin{enumerate}
    \item ERX, the proposed algorithm.
    \item RX Baseline.
    \item RT-CK-RXD \cite{chen2014real}.
    \item RX-BIL \cite{du2009fast}.
    \item LBL-AD \cite{horstrand2019novel}.  
\end{enumerate}

ERX uses a 99 line buffer to estimate the exponentially moving mean and covariance before detection starts. A momentum factor of $ \alpha = 0.1 $ is used, and the bands are reduced to 5 dimensions via sparse random projection (that is, $ d = 5 $) for all experiments. The RX Baseline algorithm serves as a benchmark for real-time anomaly detection. A rolling buffer stores the most recent 99 lines. The mean vector and covariance matrix are calculated using the entire buffer, and the RX algorithm is applied to the centre line. This process is repeated for each new line.

% \begin{figure}
%     \centering
%     \includegraphics[width=\linewidth]{Images/Method/rx_baseline.png}
%     \caption{Visualisation of the RX Baseline algorithm. This process is repeated for each new line captured by a line-scan camera in real-time.}
%     \label{rx_baseline}
% \end{figure}

RT-CK-RXD detects anomalies using a pixel-wise Woodbury matrix identity implementation. A buffer of 99 lines is used for the initial mean and covariance estimates. Just-in-time compilation via Numba \cite{lam2015numba} is also used to speed up inverse covariance updates. RX-BIL uses a line-by-line Woodbury matrix identity implementation, as well as the correlation matrix instead of the covariance matric. A 99 line buffer is also used and $\eta=50\%$ of the pixels per line are randomly discarded. Instead of using small random numbers to initialise the mean vector and correlation matrices as described in the paper, the mean and correlation of the first line are used for initialisation, substantially speeding up convergence.

LBL-AD enhances scalability using real-time PCA, offering a good comparison for ERX's dimensionality reduction via SRP. The bands are reduced to three principal components ($ d = 3 $). Numba is used to speed up the power iteration algorithm. The implementations of all algorithms are available on the GitHub repository.

\subsection{Metrics} 

The speed of each algorithm is simply measured as the average number of lines processed per second (LPS) for a given dataset. The detection performance of each algorithm is measured primarily by using the average receiver operating characteristic (ROC) curve. The ROC curve plots the true positive rate (the probability of detecting an anomaly) against the false positive rate (the probability of classifying a normal pixel as an anomaly) across all potential thresholds ($\tau$):
\begin{align}
    \text{TPR} &= \frac{\text{True Positives}}{\text{True Positives} + \text{False Negatives}} \\
    \text{FPR} &= \frac{\text{False Positives}}{\text{False Positives} + \text{True Negatives}}
\end{align}

An algorithm's performance improves as its ROC curve nears the top left of the ROC plot, indicating a good trade-off between detection and false alarms. Average ROC curves better evaluate ERX's performance over repeated experiments due to its inherent variation from sparse random projection. The variance of the model is captured by the standard deviation of the ROC curves.

The area under each ROC curve (AUC) summarises the performance in a single metric. The AUC ranges from 0 to 1, where a score of 1 indicates a perfect model, and 0.5 indicates a model that is randomly guessing. A score below 0.5 indicates that the model consistently misclassifies anomalies as background, operating contrary to the intended task. Two additional metrics are calculated for model comparison: $\text{AUC}_{\text{TD}}$ to isolate target detectability and $\text{AUC}_{\text{BS}}$ to assess background suppressibility \cite{chang20203DROC}. These metrics are slightly modified from their original form to have scores between 0 and 1:
\begin{align}
    \text{AUC}_{\text{TD}} &= \frac{\text{AUC} + \text{AUC}_{\text{TPR},\tau}}{2} \\
    \text{AUC}_{\text{BS}} &= \frac{\text{AUC} - \text{AUC}_{\text{FPR},\tau} + 1}{2}
\end{align}
\noindent
where $\text{AUC}_{\text{TPR},\tau}$ is the area under the true positive rate and thresholds curve, and $\text{AUC}_{\text{FPR},\tau}$ is the area under the false positive rate and thresholds curve.

\subsection{Experiments}

The first experiment compares the speed and scalability of all algorithms, while the second evaluates the adaptability and robustness by analysing the detection performance. An ablation study on ERX examines the impact of varying momentum and using exponentially moving statistics, and assesses the effect of dimensionality reduction using sparse random projection. 

All of the experiments are conducted on an NVIDIA Jetson Xavier NX device, a compact edge computer ideal for real-time applications onboard small platforms such as a drone. This device features 8GB of RAM, a 6-core Carmel ARMv8.2 CPU, and is selected to operate in the 20W 6-core power mode. Real-time anomaly detection is simulated in a Python environment using a sampler and a detector. The sampler sequentially feeds each line from left to right in the HSI datasets, to the detector (the algorithm being tested).

\subsubsection{Speed Test}

This experiment compares the processing speed and scalability of the detection algorithms. Each algorithm is first tested on the same randomly generated dataset of 500 pixels, 3,000 lines, and an increasing number of bands between 10 to 200. The algorithms are then tested on another generated dataset with 3,000 lines, 50 bands, and an increasing number of pixels between 100 and 1,500. Both tests are repeated five times to get the average number of lines processed per second (LPS).

\subsubsection{Detection Test}

This experiment assesses the anomaly detection performance of each algorithm on the three datasets. Each model is also tested on the flipped version of each dataset (i.e. processed from right to left instead of left to right) to further assess the adaptability of each algorithm. This is particularly relevant for the beach and synthetic datasets, which start and end with different terrains. The tests are repeated five times to obtain average results and standard deviations, indicating model variability.

\subsubsection{Ablation Study}

This study analyses how altering or removing sparse random projection and the exponentially moving mean and covariance affect ERX's performance. The first part adjusts the number of dimensions ($d$) reduced by sparse random projection: from 1 to 50 for the beach and synthetic datasets, and from 1 to 10 for the SNP dataset (since it only has 13 bands). ROC curves will be generated for all datasets, and the speed will also be assessed because of its dependence on the number of dimensions. Only AUCs and speeds are provided for the synthetic dataset, as it presents the greatest detection challenge and changing $d$ has a consistent impact on speed regardless of the dataset.

The second part of the ablation study investigates the ERX momentum parameter $\alpha$, which modifies ERX's adaptability, and detection performance based anomaly size and frequency. Different values are tested between $\alpha=1$ (using only the mean and covariance of the current line for detection) and $\alpha=1e-4$ (heavily weighing past data). This component is also eliminated by using incremental mean and covariance (IMC) updates. The implementation of the line-by-line IMC follows a batch version of Welford's algorithm for online mean and covariance updates \cite{schubert2018WelfordBatched}, which weighs all previous lines equally.

\section{Results \& Discussion} \label{sec:discussion}

\subsection{Speed Comparison} \label{sec:speed_comp}

\begin{figure}[h!]
    \centering
    \includegraphics[width=\linewidth]{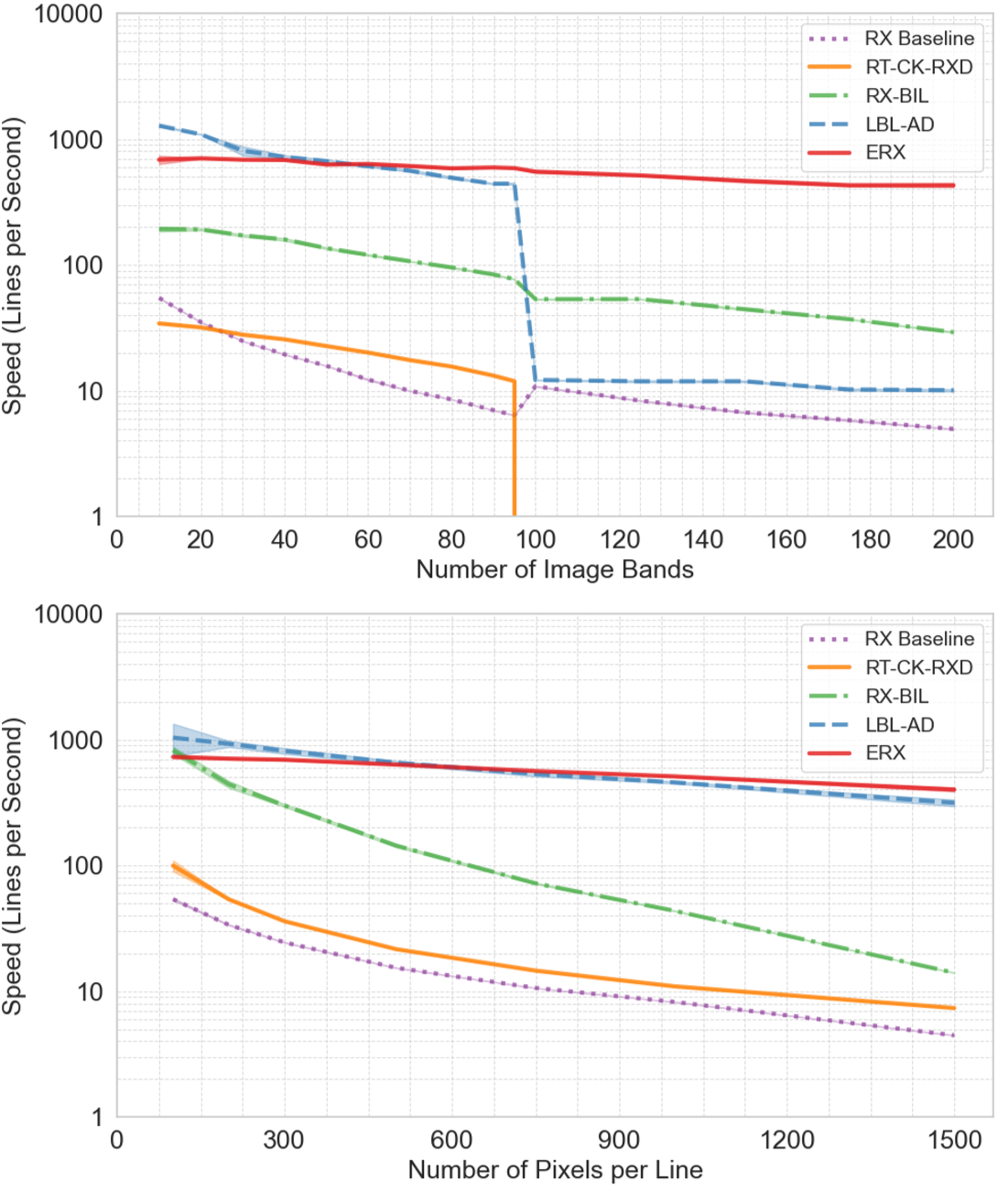}
    \caption{A comparison of the algorithms' average speed on the Jetson device across five iterations. The top graph shows speed versus the number of bands, and the bottom graph shows speed versus the number of pixels per line. The y-axis (speed) is on a log-10 scale, and the shaded areas represent the standard deviation. Note: the drastic speed change in the top graph occurs when the number of bands exceeds 95, due to the OpenBLAS library switching threading algorithms.}
    \label{fig:speed_comparison}
\end{figure}

Figure \ref{fig:speed_comparison} displays the results of the speed comparison: the top graph illustrates the speed changes with increasing bands, while the bottom graph shows the speed changes with increasing pixels per line (i.e., increasing line width). 

Initially, ERX and LBL-AD are consistently faster than RX-Baseline, RX-BIL and RT-CK-RXD as the number of bands increases. However, ERX maintains the best performance as the number of bands moves past 95. The sharp drop in LBL-AD's performance was found to occur upstream of Python, where the OpenBLAS library for linear algebra \cite{wang2013openblas} switches from single-threading to multi-threading after a dimension size of 95. This issue arises when calling the Numpy dot product, which ERX bypasses to retain higher speeds than the other algorithms. RX-BIL and RX Baseline experience slight performance changes, whereas RT-CK-RXD becomes impractical beyond 95 bands. Putting the results in the context of real-time operations, a variant of the camera that collected the beach dataset (108 band) operated at 120 lines per second with the Jetson. ERX, processing at 561 lines per second (Table \ref{tab:detection_results}), is the only algorithm fast enough for real-time anomaly detection. It is also the fastest algorithm for the synthetic dataset and the second fastest for the SNP dataset.

\begin{figure*}[t]
    \centering
    \begin{minipage}{\linewidth}
        % Row 1 for Beach dataset
        \setcounter{subfigure}{0}
        \begin{subfigure}{0.32\linewidth}
            \centering
            \includegraphics[width=\linewidth]{Images/Method/beach_gt.png}
            \subcaption{Beach dataset ground truth}
        \end{subfigure}
        \begin{subfigure}{0.32\linewidth}
            \centering
            \includegraphics[width=\linewidth]{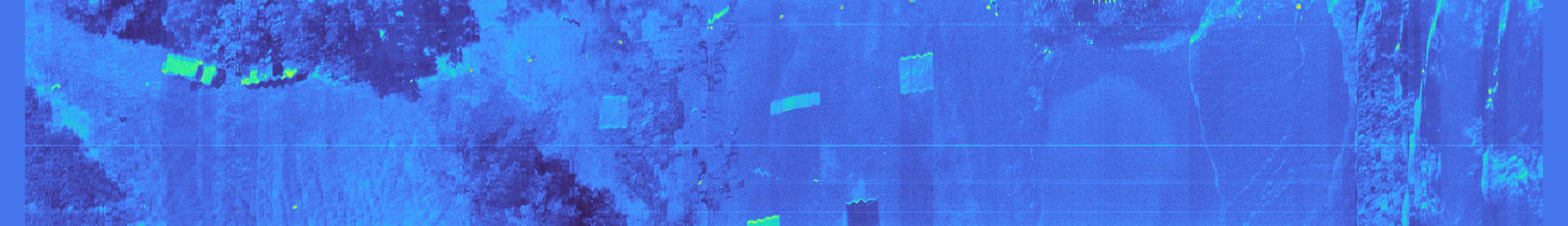}
            \subcaption{RX Baseline}
        \end{subfigure}
        \begin{subfigure}{0.32\linewidth}
            \centering
            \includegraphics[width=\linewidth]{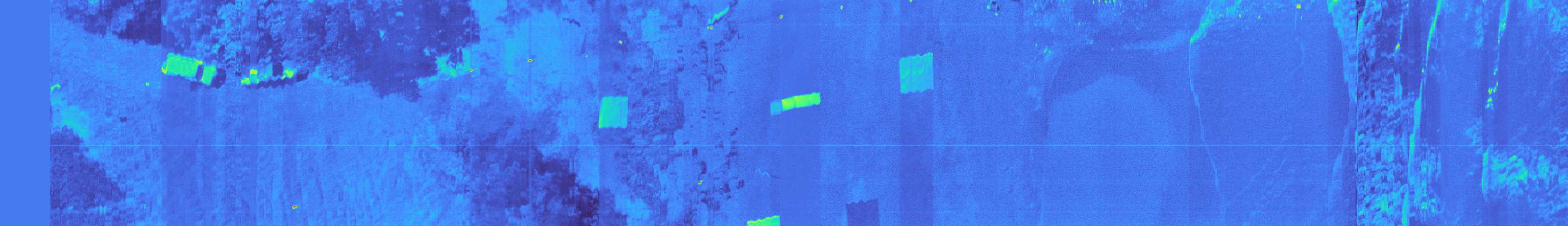}
            \subcaption{RT-CK-RXD}
        \end{subfigure}
        \\ % Row 2 for Beach dataset
        \begin{subfigure}{0.32\linewidth}
            \centering
            \includegraphics[width=\linewidth]{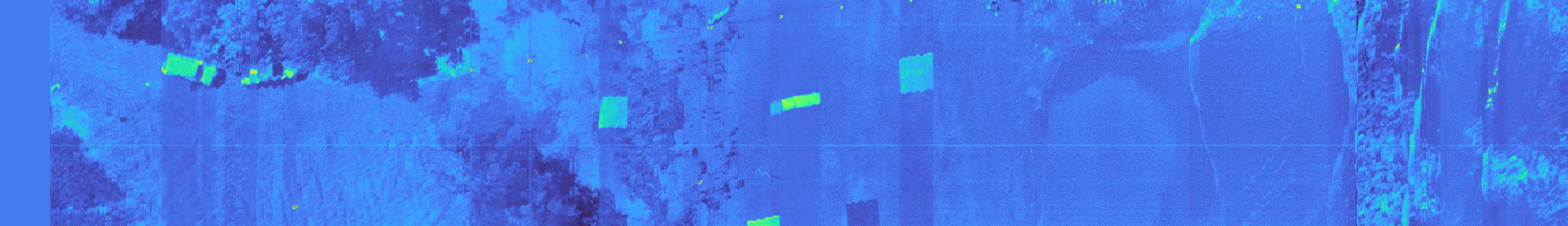}
            \subcaption{RX-BIL}
        \end{subfigure}
        \begin{subfigure}{0.32\linewidth}
            \centering
            \includegraphics[width=\linewidth]{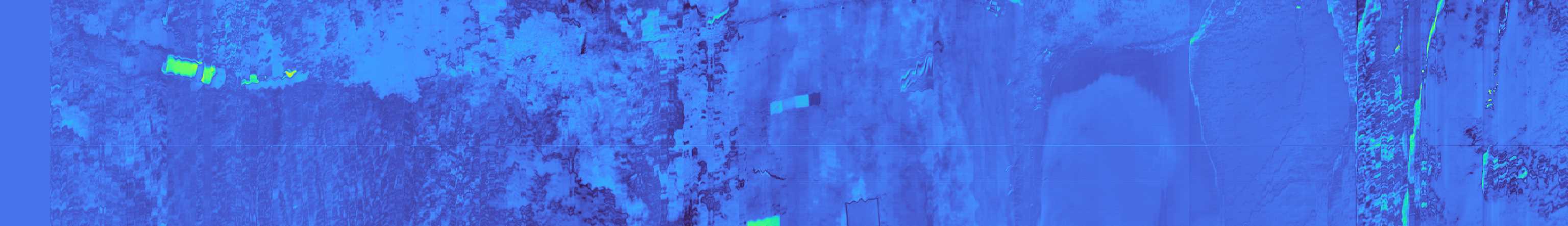}
            \subcaption{LBL-AD}
        \end{subfigure}
        \begin{subfigure}{0.32\linewidth}
            \centering
            \includegraphics[width=\linewidth]{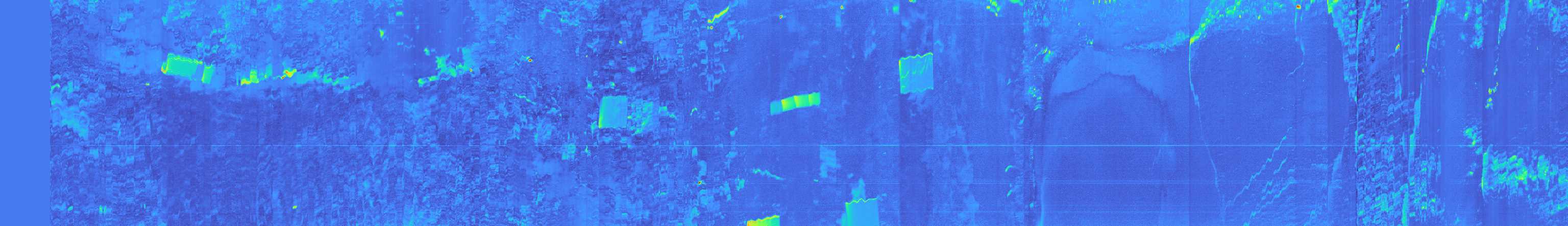}
            \subcaption{ERX}
        \end{subfigure}
    \end{minipage}

    \vspace{0.5cm}

    \begin{minipage}{\linewidth}
        % Row 1 for Synthetic dataset
        \setcounter{subfigure}{0}
        \begin{subfigure}{0.32\linewidth}
            \centering
            \includegraphics[width=\linewidth]{Images/Method/synthetic_gt.png}
            \subcaption{Synthetic dataset ground truth}
        \end{subfigure}
        \begin{subfigure}{0.32\linewidth}
            \centering
            \includegraphics[width=\linewidth]{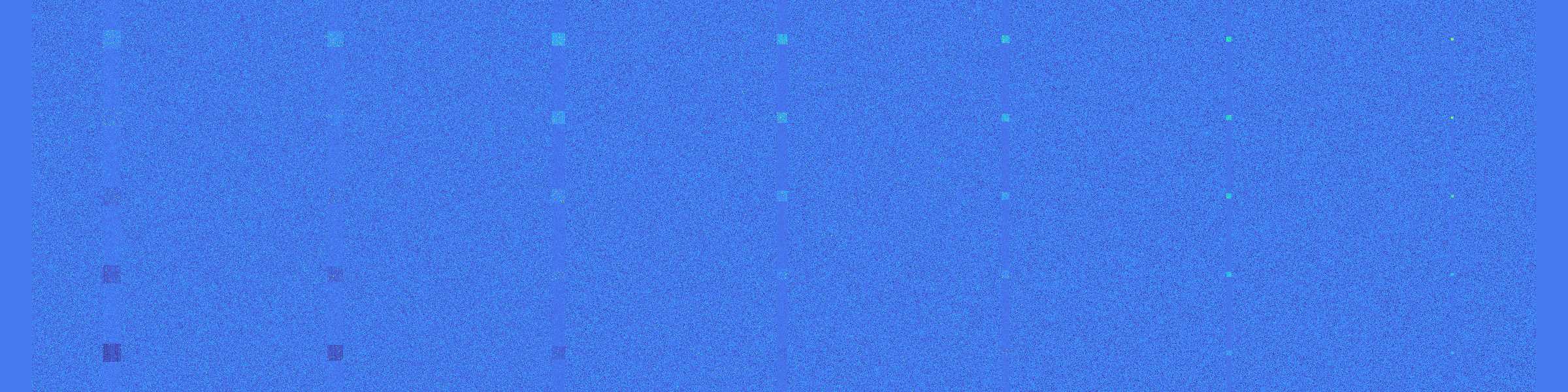}
            \subcaption{RX Baseline}
        \end{subfigure}
        \begin{subfigure}{0.32\linewidth}
            \centering
            \includegraphics[width=\linewidth]{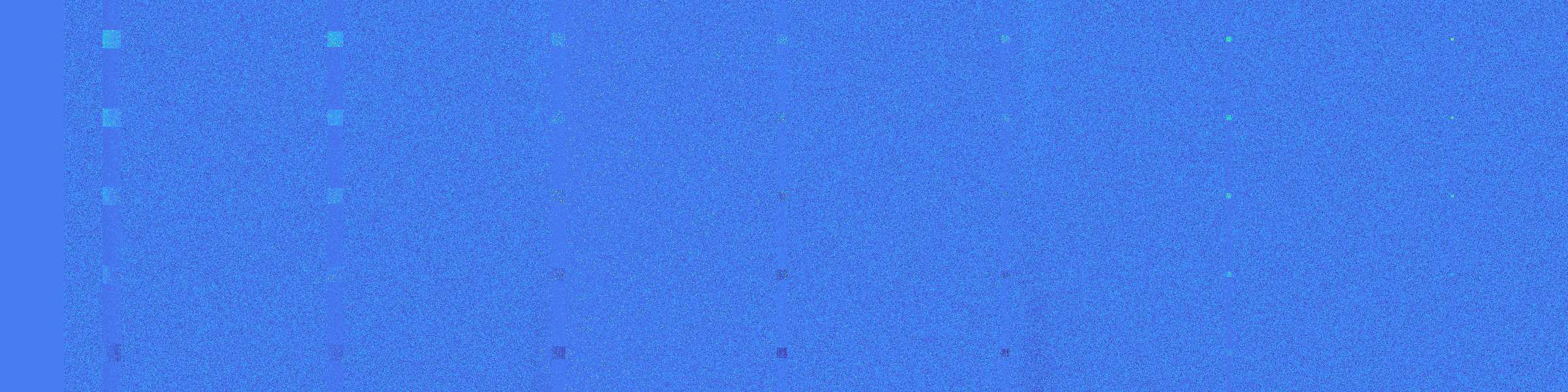}
            \subcaption{RT-CK-RXD}
        \end{subfigure}
        \\ % Row 2 for Synthetic dataset
        \begin{subfigure}{0.32\linewidth}
            \centering
            \includegraphics[width=\linewidth]{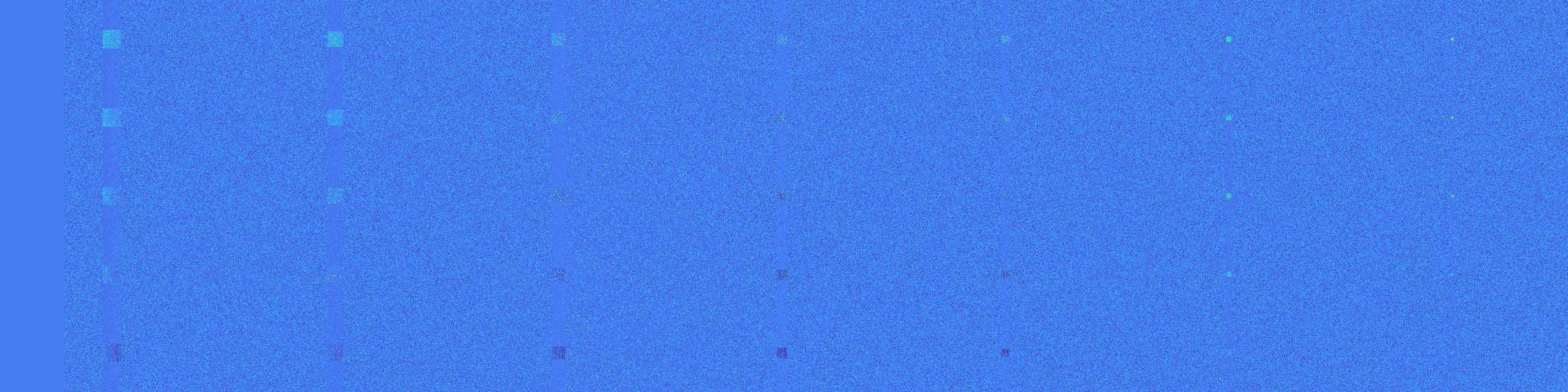}
            \subcaption{RX-BIL}
        \end{subfigure}
        \begin{subfigure}{0.32\linewidth}
            \centering
            \includegraphics[width=\linewidth]{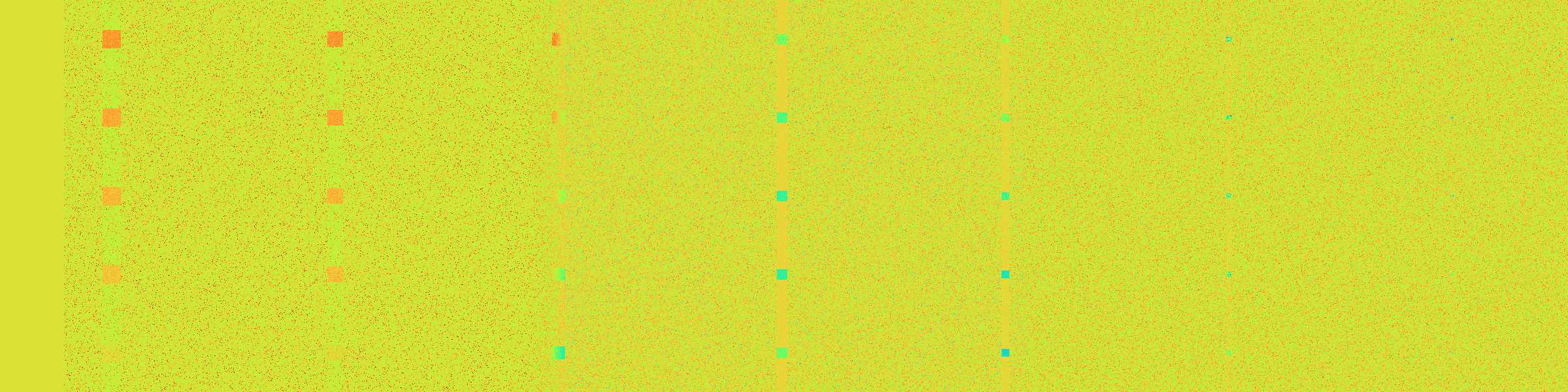}
            \subcaption{LBL-AD}
        \end{subfigure}
        \begin{subfigure}{0.32\linewidth}
            \centering
            \includegraphics[width=\linewidth]{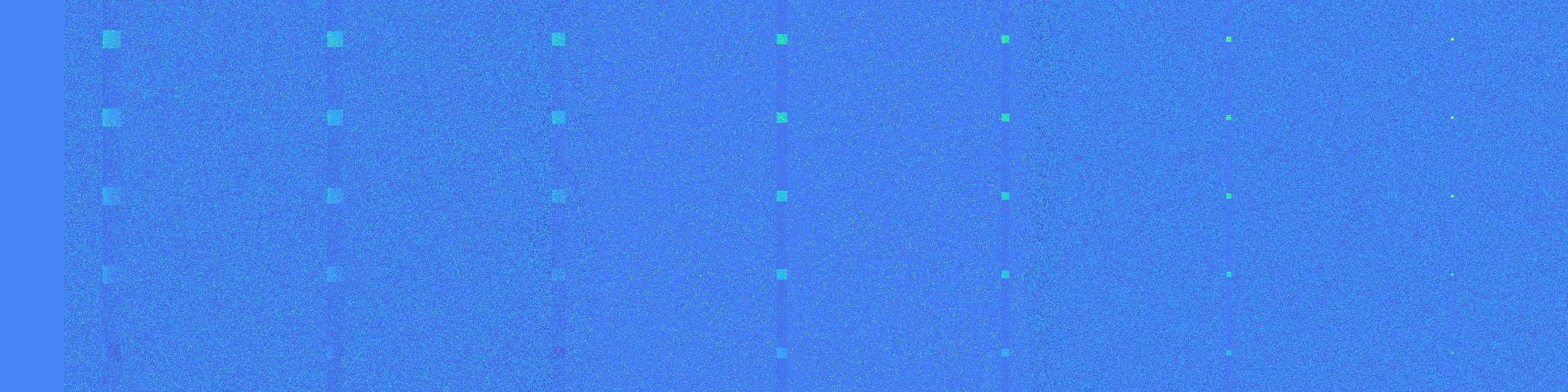}
            \subcaption{ERX}
        \end{subfigure}
    \end{minipage}

    \vspace{0.5cm}

    \begin{minipage}{\linewidth}
        % Row 1 for SNP dataset
        \setcounter{subfigure}{0}
        \begin{subfigure}{0.32\linewidth}
            \centering
            \includegraphics[width=\linewidth]{Images/Method/snp_gt.png}
            \subcaption{SNP dataset ground truth}
        \end{subfigure}
        \begin{subfigure}{0.32\linewidth}
            \centering
            \includegraphics[width=\linewidth]{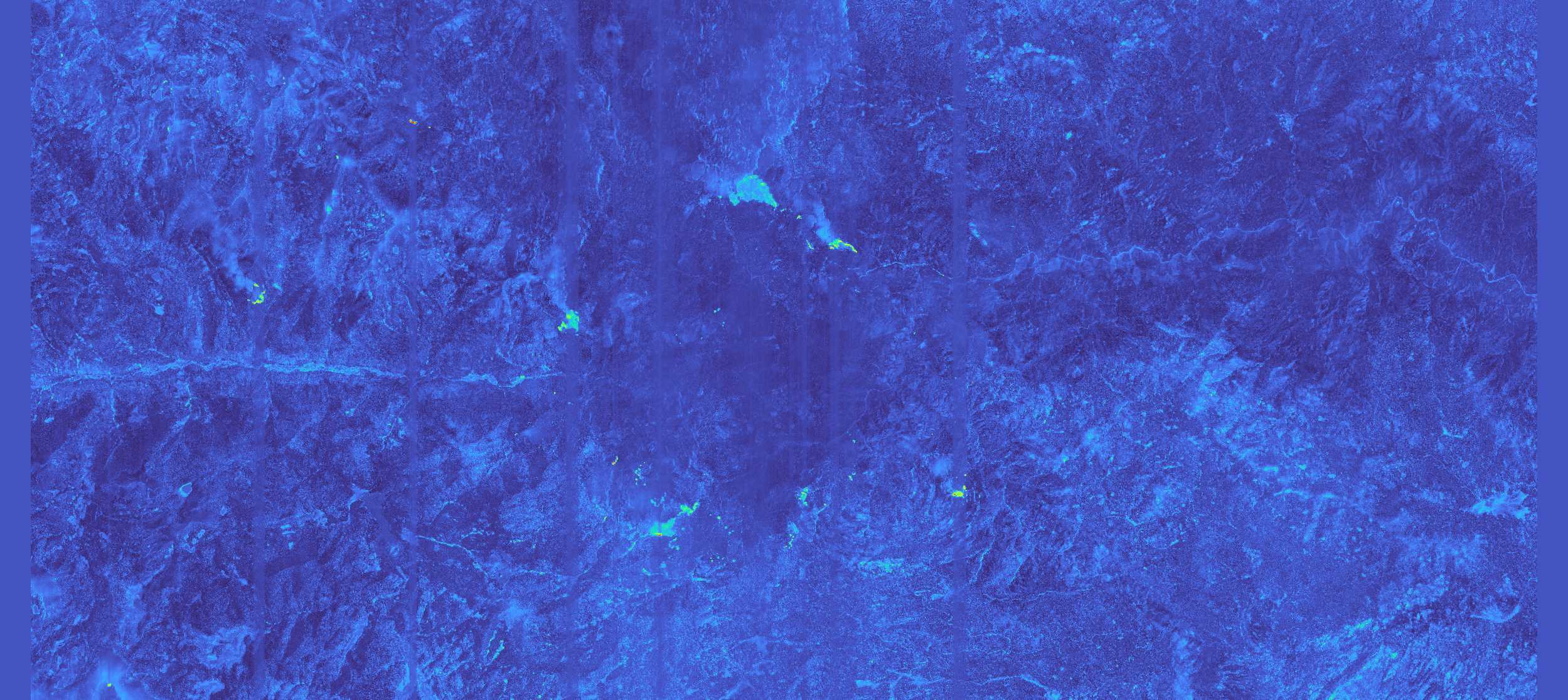}
            \subcaption{RX Baseline}
        \end{subfigure}
        \begin{subfigure}{0.32\linewidth}
            \centering
            \includegraphics[width=\linewidth]{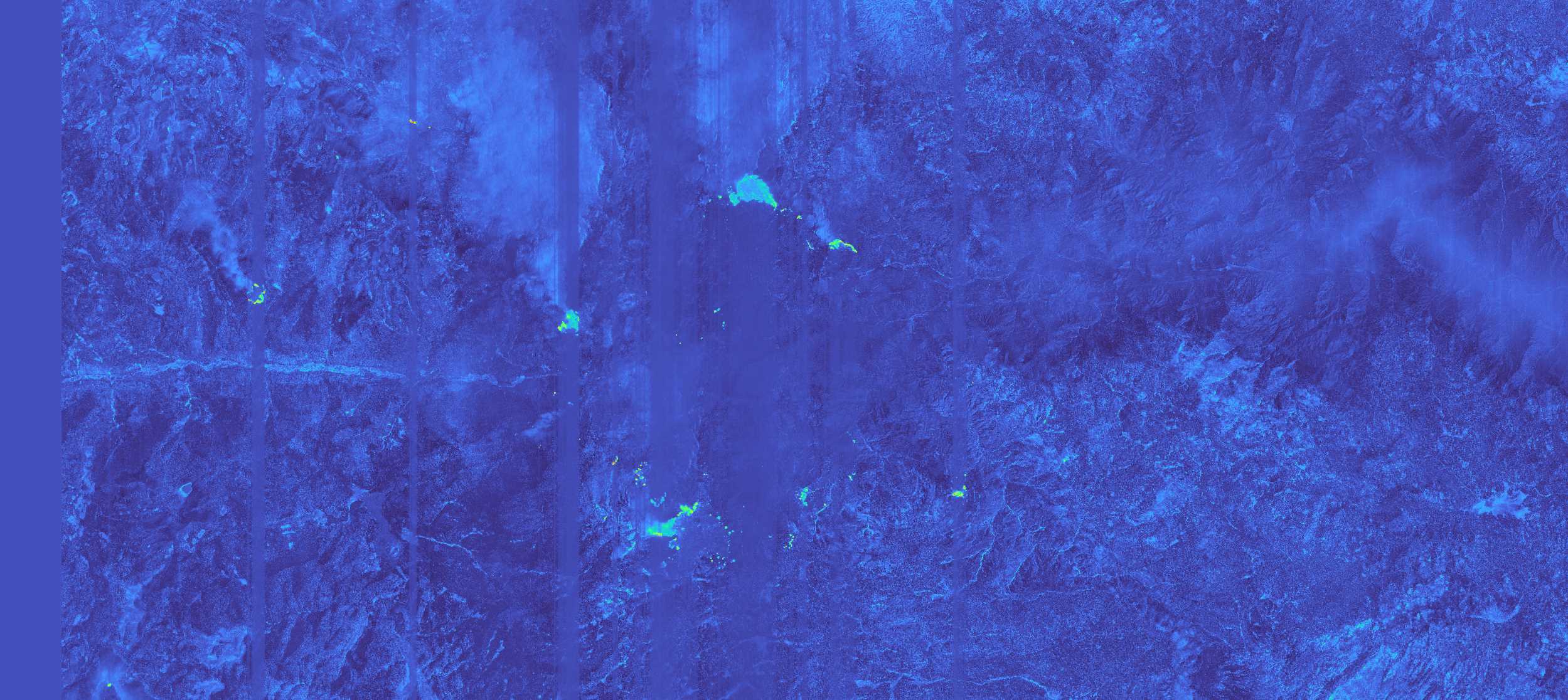}
            \subcaption{RT-CK-RXD}
        \end{subfigure}
        \\ % Row 2 for SNP dataset
        \begin{subfigure}{0.32\linewidth}
            \centering
            \includegraphics[width=\linewidth]{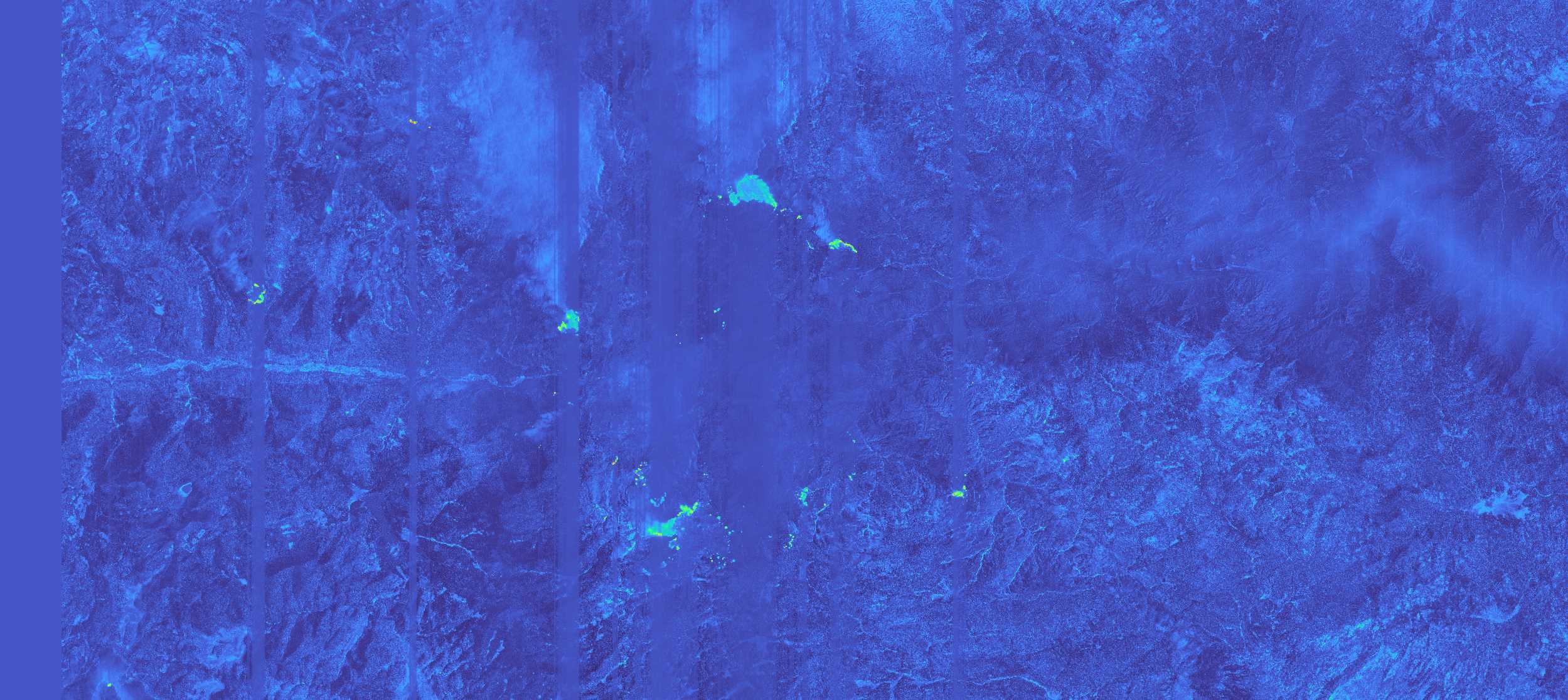}
            \subcaption{RX-BIL}
        \end{subfigure}
        \begin{subfigure}{0.32\linewidth}
            \centering
            \includegraphics[width=\linewidth]{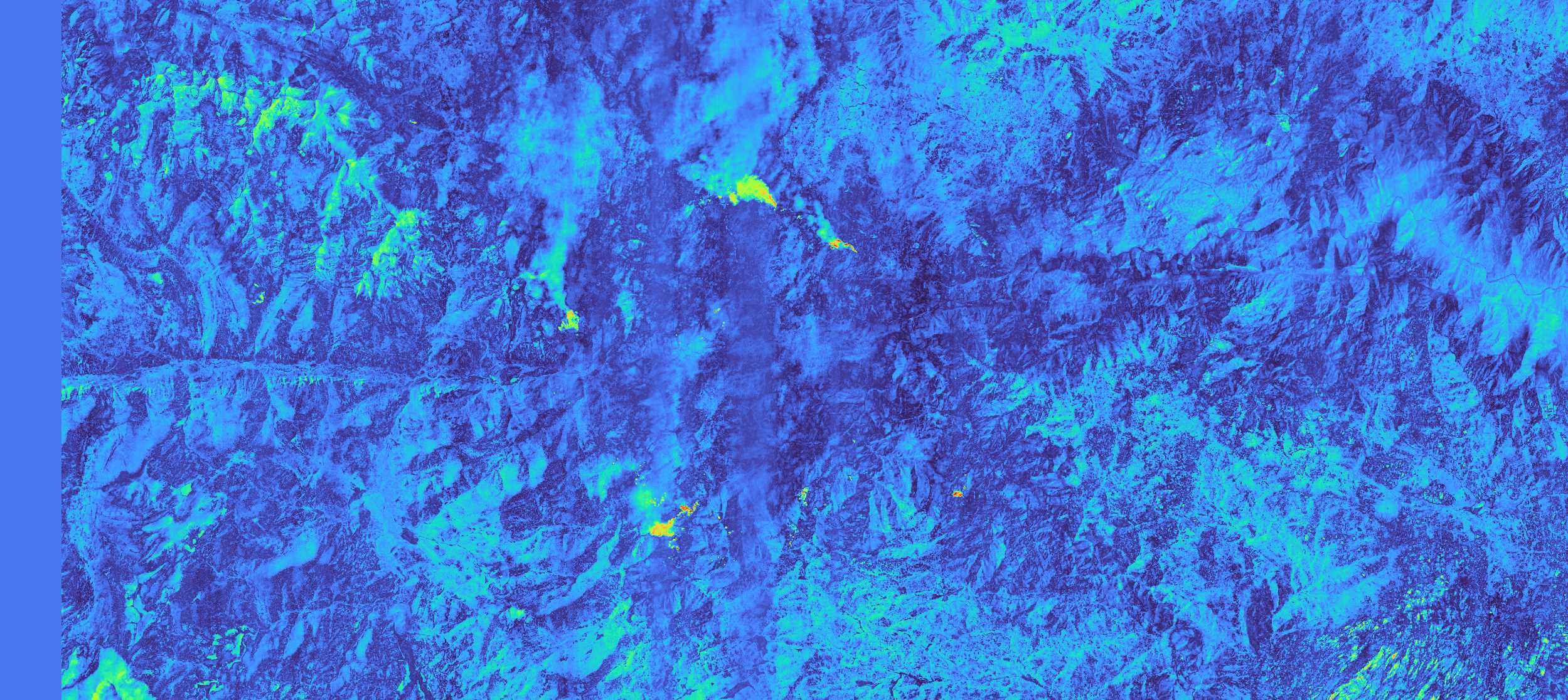}
            \subcaption{LBL-AD}
        \end{subfigure}
        \begin{subfigure}{0.32\linewidth}
            \centering
            \includegraphics[width=\linewidth]{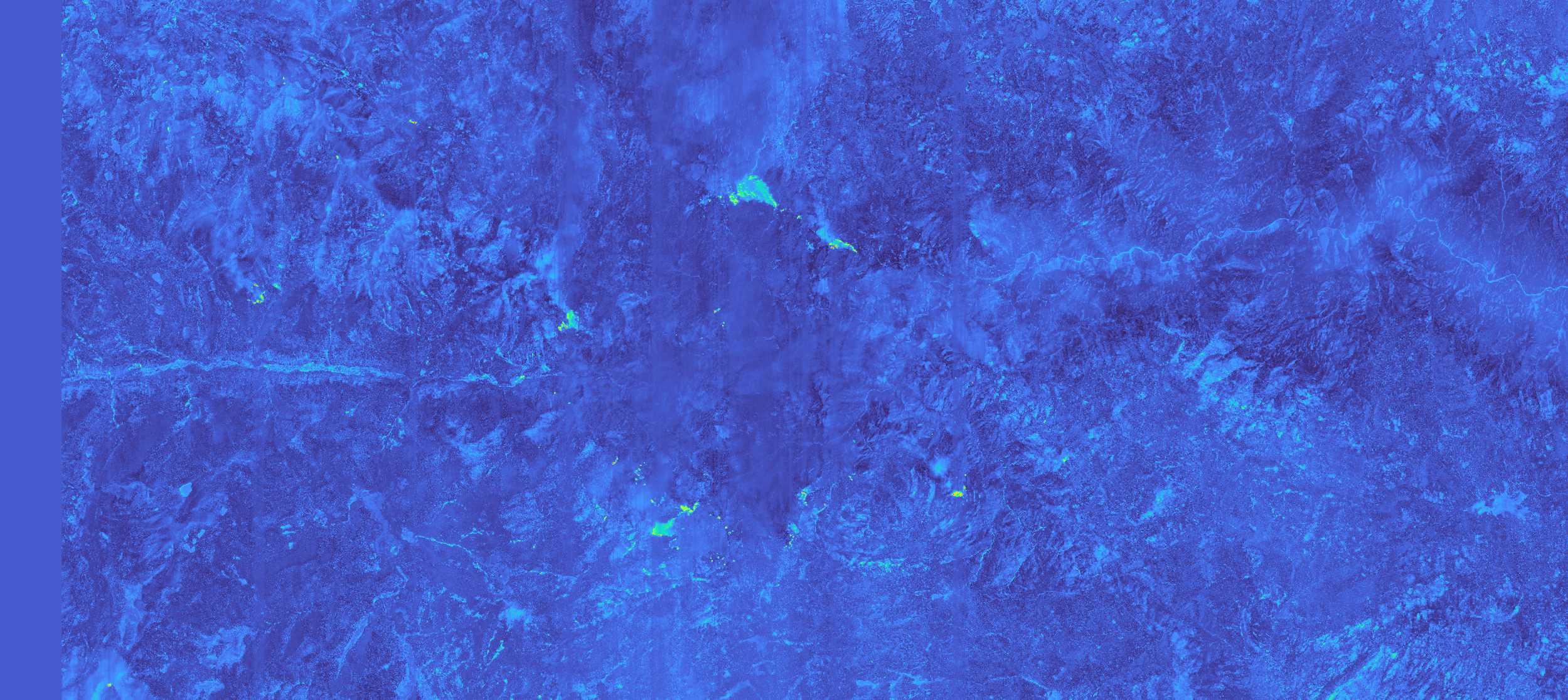}
            \subcaption{ERX}
        \end{subfigure}
    \end{minipage}

     \begin{minipage}{0.55\textwidth}
        \centering
        \includegraphics[width=\textwidth]{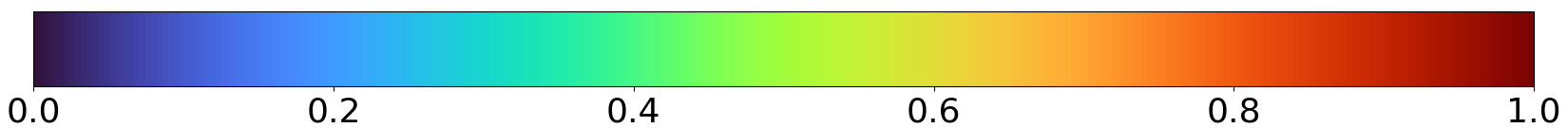}
    \end{minipage}%
    
    \caption{Anomaly heatmaps from each model for beach (top), synthetic (middle), and SNP (bottom) datasets. The top 10 pixels of each line are excluded from all beach heatmaps due to large radiometric distortions skewing the colour bar.}
    \label{fig:detection_maps}
\end{figure*}

Figure \ref{fig:speed_comparison} shows the RX-Baseline, RX-BIL, and RT-CK-RXD speeds rapidly decrease with increasing line width, while ERX and LBL-AD remain fast. There is a minor initial variation for LBL-AD attributed to processing noise on the Jetson, given that performance remains generally consistent for the remainder of the graph. In general, ERX is demonstrated as the most stable and scalable algorithm in this experiment.

\subsection{Detection Comparison}

\begin{figure*}[p]
    \centering
    \begin{minipage}{0.32\textwidth}
        \centering
        \includegraphics[width=\textwidth]{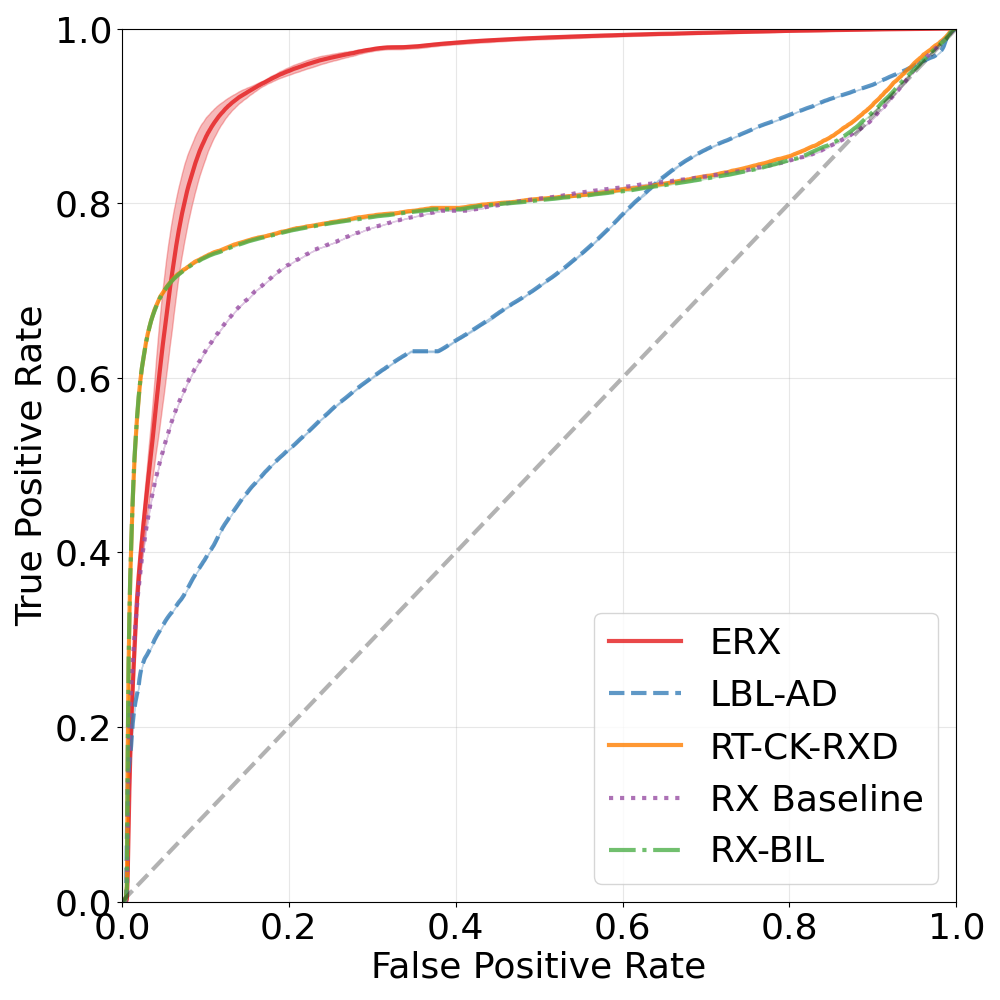}
        \subcaption{Beach}
        \label{roc:detection_beach}
    \end{minipage}%
    \hfill
    \begin{minipage}{0.32\textwidth}
        \centering
        \includegraphics[width=\textwidth]{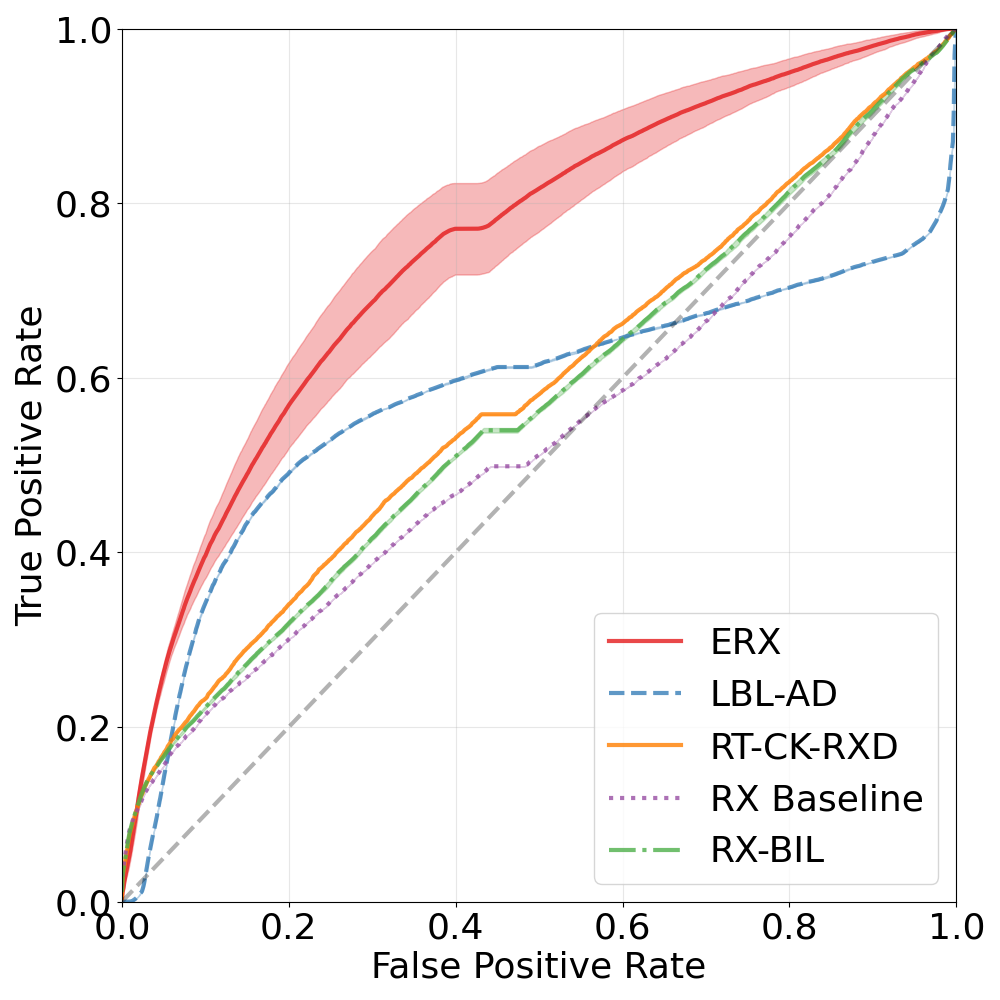}
        \subcaption{Synthetic}
        \label{roc:detection_synthetic}
    \end{minipage}%
    \hfill
    \begin{minipage}{0.32\textwidth}
        \centering
        \includegraphics[width=\textwidth]{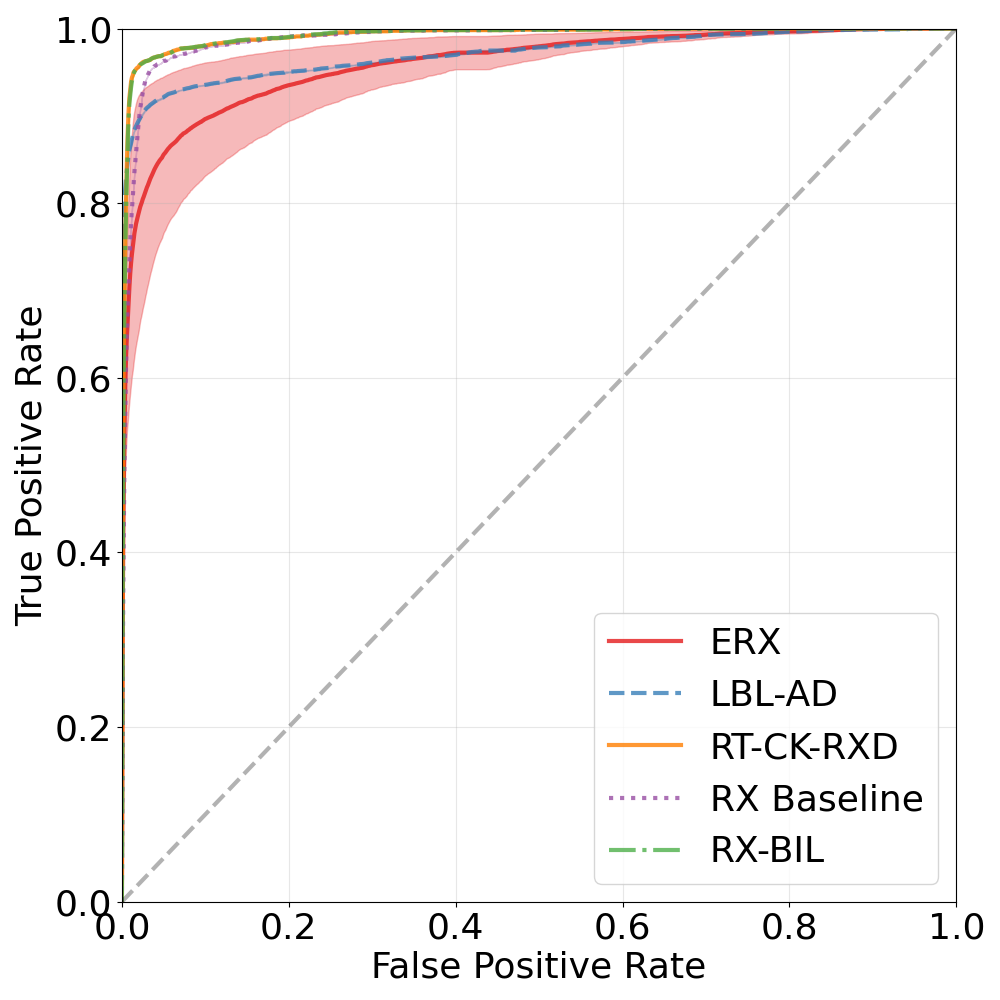}
        \subcaption{SNP}
        \label{roc:detection_snp}
    \end{minipage} 

    \begin{minipage}{0.32\textwidth}
        \centering
        \includegraphics[width=\textwidth]{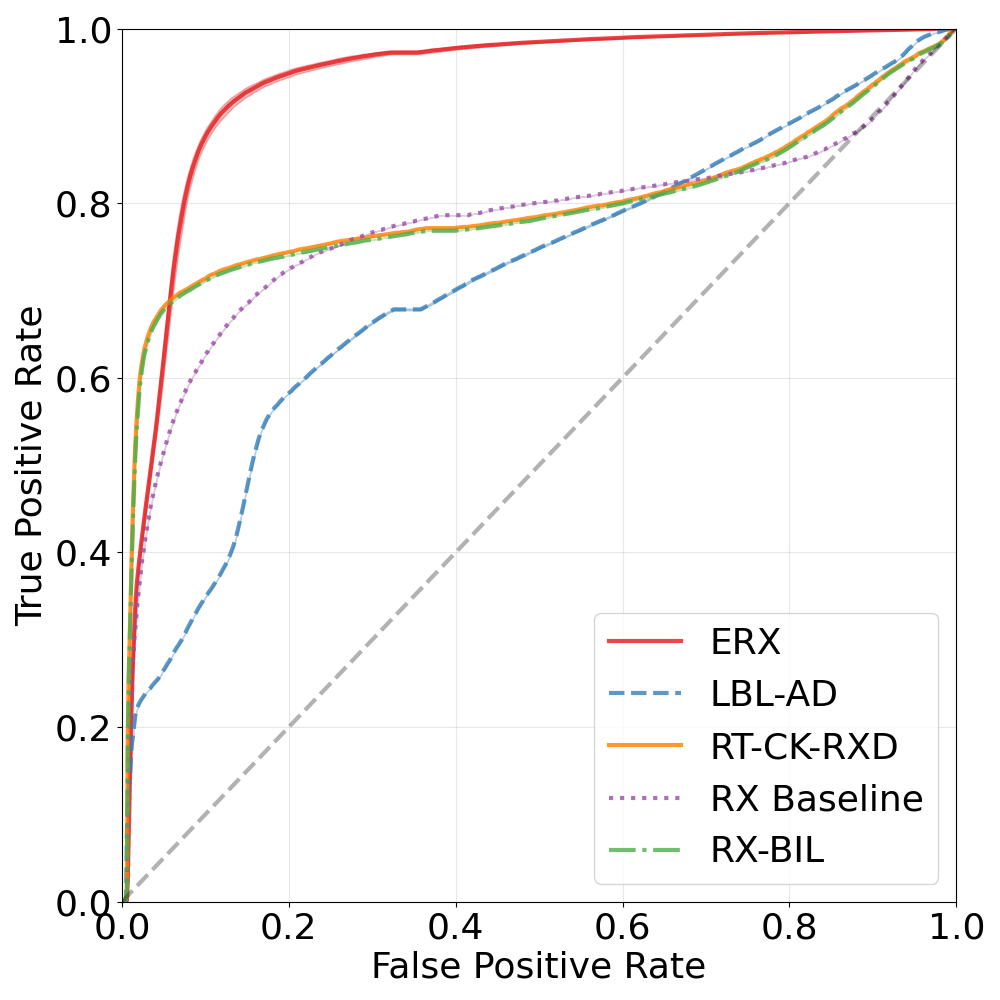}
        \subcaption{Beach (Flipped)}
        \label{roc:detection_beach_flipped}
    \end{minipage}%
    \hfill
    \begin{minipage}{0.32\textwidth}
        \centering
        \includegraphics[width=\textwidth]{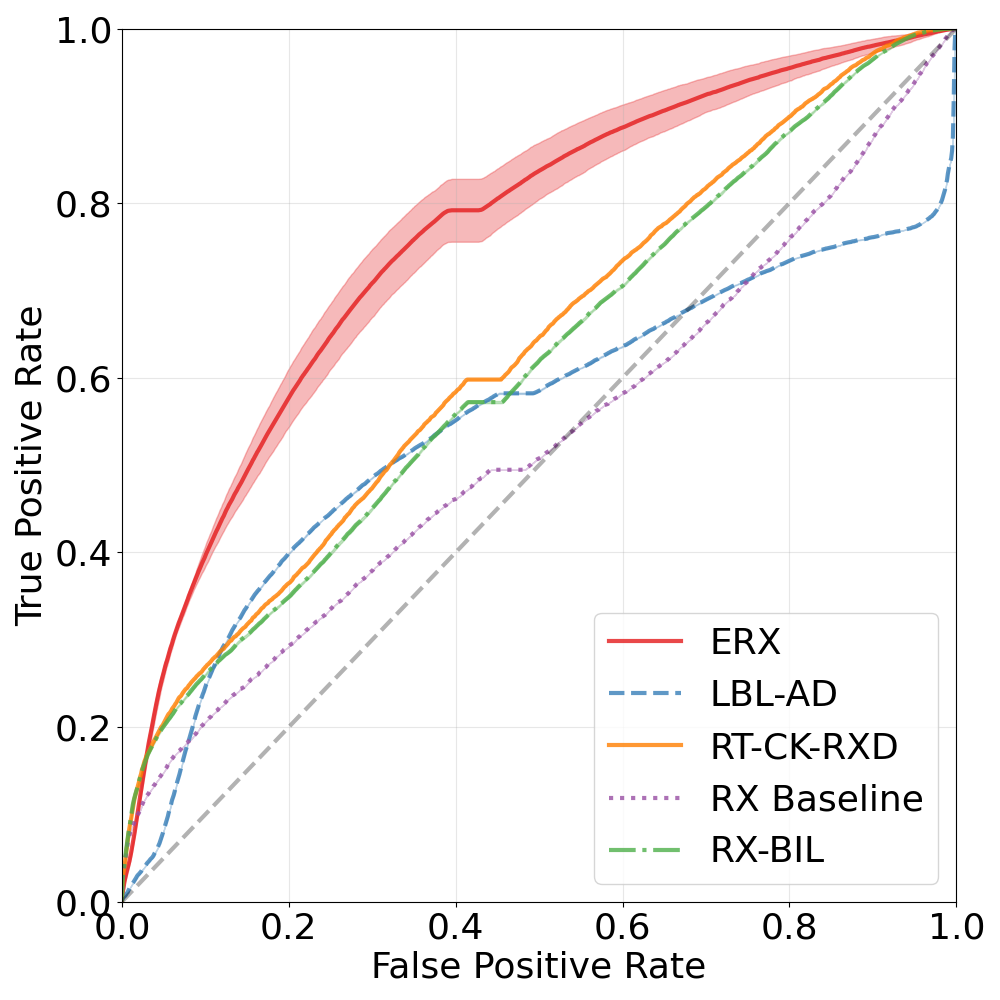}
        \subcaption{Synthetic (Flipped)}
        \label{roc:detection_synthetic_flipped}
    \end{minipage}%
    \hfill
    \begin{minipage}{0.32\textwidth}
        \centering
        \includegraphics[width=\textwidth]{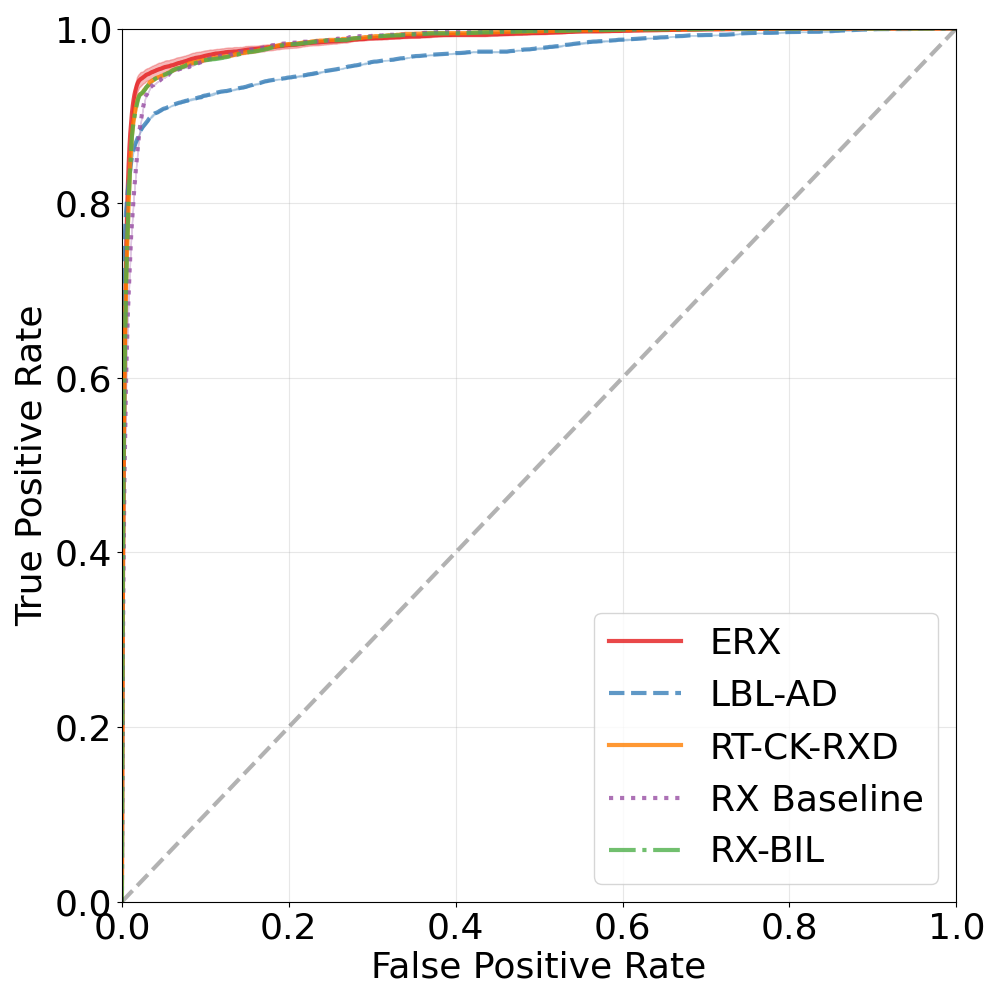}
        \subcaption{SNP (Flipped)}
        \label{roc:detection_SNP_flipped}
    \end{minipage} 
    \caption{Mean ROC curves for all models across each dataset (both normal and flipped). The shaded area shows the standard deviation across five iterations.}
    \label{fig:roc_detection}
\end{figure*}

Figure \ref{fig:detection_maps} displays the models' anomaly heatmaps for the original datasets. Figure \ref{fig:roc_detection} shows the mean ROC curves of all models for the original datasets and their flipped versions. The aggregated performance metrics are presented in Table \ref{tab:detection_results}.

\begin{table*}[p]
\caption{Detection Results Across Five Iterations}
\centering
%\subcaption{}
\label{tab:beach_detection}
\begin{tabular}{@{}c|ccc|c|ccc@{}}
\toprule
            & \multicolumn{3}{c|}{\textbf{Beach}}                                                           & \textbf{Speed} & \multicolumn{3}{c}{\textbf{Beach (Flipped)}}                                                           \\
Model       & AUC $\pm$ SD           & $\text{AUC}_{\text{TD}}$ $\pm$ SD & $\text{AUC}_{\text{BS}}$ $\pm$ SD & LPS            & AUC $\pm$ SD           & $\text{AUC}_{\text{TD}}$ $\pm$ SD & $\text{AUC}_{\text{BS}}$ $\pm$ SD \\ \midrule
RX Baseline & $0.778 \pm 0.000$      & $0.821 \pm 0.000$                 & $0.482 \pm 0.000$                 & $12$           & $0.774 \pm 0.000$      & $0.820 \pm 0.000$                 & $0.480 \pm 0.000$                 \\
RT-CK-RXD   & $0.805 \pm 0.000$      & $0.826 \pm 0.000$                 & $0.515 \pm 0.000$                 & $<1$           & $0.795 \pm 0.000$      & $0.833 \pm 0.000$                 & $0.498 \pm 0.000$                 \\
RX-BIL      & $0.803 \pm 0.000$      & $0.828\pm 0.000$                  & $0.510 \pm 0.001$                 & $62$           & $0.792 \pm 0.000$      & $0.833 \pm 0.000$                 & $0.494 \pm 0.001$                 \\
LBL-AD      & $0.695 \pm 0.000$      & $0.792 \pm 0.000$                 & $0.416 \pm 0.000$                 & $12$           & $0.710 \pm 0.000$      & $0.798 \pm 0.000$                 & $0.428 \pm 0.000$                 \\
ERX         & $\bm{0.941 \pm 0.005}$ & $\bm{0.890 \pm 0.006}$            & $\bm{0.659 \pm 0.017}$            & $\bm{561}$     & $\bm{0.938 \pm 0.002}$ & $\bm{0.896 \pm 0.002}$            & $\bm{0.635 \pm 0.008}$            \\ \bottomrule
\end{tabular}
\vspace{0.25cm}
\centering
%\subcaption{}
\label{tab:synthetic_detection}
\begin{tabular}{@{}c|ccc|c|ccc@{}}
\toprule
            & \multicolumn{3}{c|}{\textbf{Synthetic}}                                                      & \textbf{Speed} & \multicolumn{3}{c}{\textbf{Synthetic (Flipped)}}                                                     \\
Model       & AUC $\pm$ SD      & $\text{AUC}_{\text{TD}}$ $\pm$ SD & $\text{AUC}_{\text{BS}}$ $\pm$ SD & LPS            & AUC $\pm$ SD      & $\text{AUC}_{\text{TD}}$ $\pm$ SD & $\text{AUC}_{\text{BS}}$ $\pm$ SD \\ \midrule
RX Baseline & $0.531 \pm 0.000$ & $0.699 \pm 0.000$                 & $0.336 \pm 0.000$                 & $4$            & $0.527 \pm 0.000$ & $0.696 \pm 0.000$                 & $0.334 \pm 0.000$                 \\
RT-CK-RXD   & $0.581 \pm 0.000$ & $0.708 \pm 0.000$                 & $0.379 \pm 0.000$                 & $10$           & $0.632 \pm 0.000$ & $0.695 \pm 0.000$                 & $0.457 \pm 0.000$                 \\
RX-BIL      & $0.566 \pm 0.002$ & $0.714 \pm 0.001$                 & $0.357 \pm 0.001$                 & $64$           & $0.615 \pm 0.001$ & $0.724 \pm 0.002$                 & $0.400 \pm 0.002$                 \\
LBL-AD      & $0.576 \pm 0.000$ & $0.509 \pm 0.001$                 & $\bm{0.563 \pm 0.000}$                 & $361$          & $0.550 \pm 0.000$ & $0.611 \pm 0.000$                 & $0.433 \pm 0.000$                 \\
ERX         & $\bm{0.751 \pm 0.032}$ & $\bm{0.807 \pm 0.019}$                 & $0.463 \pm 0.016$                 & $\bm{508}$     & $\bm{0.760 \pm 0.022}$ & $\bm{0.764 \pm 0.023}$                 & $\bm{0.514 \pm 0.004}$                 \\ \bottomrule
\end{tabular}
\vspace{0.25cm}
\centering
%\subcaption{}
\label{tab:snp_detection}
\begin{tabular}{@{}c|ccc|c|ccc@{}}
\toprule
            & \multicolumn{3}{c|}{\textbf{SNP}}                                                           & \textbf{Speed} & \multicolumn{3}{c}{\textbf{SNP (Flipped)}}                                                           \\
Model       & AUC $\pm$ SD           & $\text{AUC}_{\text{TD}}$ $\pm$ SD & $\text{AUC}_{\text{BS}}$ $\pm$ SD & LPS            & AUC $\pm$ SD           & $\text{AUC}_{\text{TD}}$ $\pm$ SD & $\text{AUC}_{\text{BS}}$ $\pm$ SD \\ \midrule
RX Baseline & $0.988 \pm 0.000$      & $\bm{0.871 \pm 0.000}$            & $\bm{0.880 \pm 0.000}$            & $38$           & $0.984 \pm 0.000$      & $0.870 \pm 0.000$                 & $\bm{0.860 \pm 0.000}$            \\
RT-CK-RXD   & $\bm{0.992 \pm 0.000}$ & $\bm{0.871 \pm 0.000}$            & $0.864 \pm 0.000$                 & $15$           & $\bm{0.986 \pm 0.000}$ & $0.883 \pm 0.000$                 & $0.815 \pm 0.000$                 \\
RX-BIL      & $\bm{0.992 \pm 0.000}$ & $0.869\pm 0.000$                  & $0.857 \pm 0.001$                 & $35$           & $\bm{0.986 \pm 0.000}$ & $\bm{0.887 \pm 0.001}$            & $0.803 \pm 0.001$                 \\
LBL-AD      & $0.971 \pm 0.000$      & $0.818 \pm 0.003$                 & $0.814 \pm 0.005$                 & $\bm{787}$     & $0.968 \pm 0.000$      & $0.835 \pm 0.002$                 & $0.779 \pm 0.004$                 \\
ERX         & $0.959 \pm 0.024$      & $0.866 \pm 0.006$                 & $0.754 \pm 0.055$                 & $561$          & $\bm{0.986 \pm 0.002}$ & $0.871 \pm 0.005$                 & $0.809 \pm 0.009$                 \\ \bottomrule
\end{tabular}
\label{tab:detection_results}
\end{table*}

Figure \ref{fig:roc_detection} shows that ERX is the overall top performer, excelling on the beach and synthetic datasets and remaining competitive on the SNP dataset, demonstrating its adaptability and robustness. ERX's ROC curves are much more desirable, vastly outperforming the other algorithms on the HSI datasets even with model variation. Figure \ref{fig:detection_maps} shows that ERX uniquely highlights the black tarp in the beach dataset. Some brighter patches of vegetation and water areas in ERX's heatmap implies higher sensitivity to local outliers. This reflects an adaptive trade-off, which loses some prior information due to the exponential moving mean and covariance. RX-BIL and RT-CK-RXD effectively highlight most targets, but LBL-AD misses most, detecting only a tarp and cars, indicating a loss of essential spectral information in the real-time PCA technique. 

Table \ref{tab:detection_results} further highlights the strong performance of ERX on the beach dataset, excelling across all detection metrics and in speed. ERX achieves high mean AUCs for both normal and flipped images, demonstrating its adaptability to changing scenery and different initial backgrounds. The larger margin between ERX and other algorithms' $\text{AUC}_{\text{BS}}$ score highlights its primary strength in reducing false alarms, suggesting its adaptive methodology excels at background suppression. RX-BIL and RT-CK-RXD detect well but are too slow for real-time use. RT-CK-RXD is especially slow on the Jetson due to the OpenBLAS threading limitations and its pixel-wise approach. ERX is over nine times faster than RX-BIL, the next-fastest algorithm. ERX's scalability, adaptability, and robustness are demonstrated given the high-dimensionality, the changing scenery, and the geometric distortions.

ERX also achieves the best results on the synthetic dataset. ERX's adaptability is evident with its high mean AUC between normal and flipped images, unlike RX-BIL and RT-CK-RXD which show significant changes. In the flipped dataset the smaller anomalies are processed first, contributing less to initial background estimates and enhancing background-anomaly separation for these two algorithms. All algorithms struggle to perform well on the this dataset given that most have AUC scores near 0.5, which is only slightly better than random guessing. It is challenging due to a higher proportion of anomalous pixels, complicating anomaly-background separation compared to the beach and SNP datasets. The higher standard deviations for ERX suggest that the five dimensions ($d=5$) used in sparse random projection do not fully capture all the information needed to consistently separate anomalies.

The high frequency of anomalies affects the RX Baseline and LBL-AD algorithms the most, as their ROC curves fall below the random model at lower thresholds (top right of Figures \ref{roc:detection_synthetic} and \ref{roc:detection_synthetic_flipped}). The impact is visualised in the anomaly heatmaps, where LBL-AD mistakenly detects the background as anomalous. Overall, ERX performs well and Figure \ref{fig:detection_maps} shows that it detects most panel structures despite a generally low mean AUC. ERX's consistently higher $\text{AUC}_{\text{TD}}$ scores support this, but the relatively low $\text{AUC}_{\text{BS}}$ scores for all algorithms highlight the general challenge of background suppression in this dataset. Object-based detection methods that factor in the spatial structure of neighbouring anomalies could be more practical than pixel-based approaches.

ERX has the second fastest speed for the Sequoia National Park dataset but slightly underperforms the other algorithms for mean AUC (Table \ref{tab:detection_results}). LBL-AD is the fastest on this lower-dimensional dataset, which is consistent with the findings in the speed comparison. All algorithms perform well on this dataset, which can be attributed to the stable background with minimal change in scenery. Visually, LBL-AD effectively highlights anomalies but exhibits a brighter background (Figure \ref{fig:detection_maps}), indicating sensitivity to noise. This aligns with the hard assumption of a homogeneous background. RX Baseline, RT-CK-RXD, and RX-BIL separate anomalies well, while ERX generally performs well but struggles slightly with background suppression. This can be seen by the brighter spots in the anomaly heatmap (similar to the beach dataset) and the lower $\text{AUC}_{\text{BS}}$ score. Reducing the momentum for slower updates should reduce this noise.

ERX has a larger model variance for the SNP dataset compared to the flipped counterpart and the other datasets, shown in Figure \ref{roc:detection_snp} and by the AUC standard deviation (Table \ref{tab:detection_results}). The reason is unclear, but the small number of iterations (five) could be the cause. More iterations could stabilise model performance. However, ERX still maintains strong absolute detection and a high speed, achieving AUCs above $0.959$ on both normal and flipped datasets.

\newpage

\subsection{Ablation Study} \label{subsec:ablation_study}

\subsubsection{Sparse Random Projection}

\begin{figure*}[t]
    \centering
    \begin{minipage}{0.32\textwidth}
        \centering
        \includegraphics[width=\textwidth]{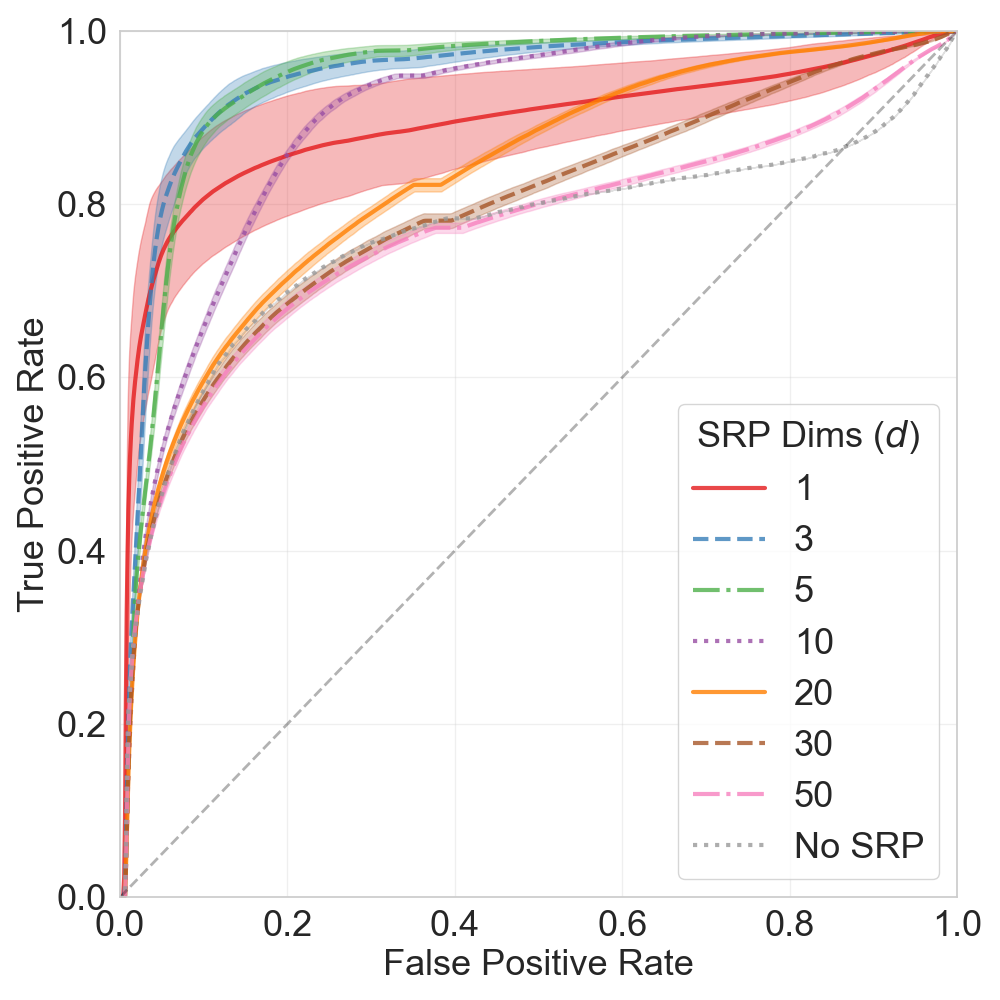}
        \subcaption{Beach}
    \end{minipage}%
    \hfill
    \begin{minipage}{0.32\textwidth}
        \centering
        \includegraphics[width=\textwidth]{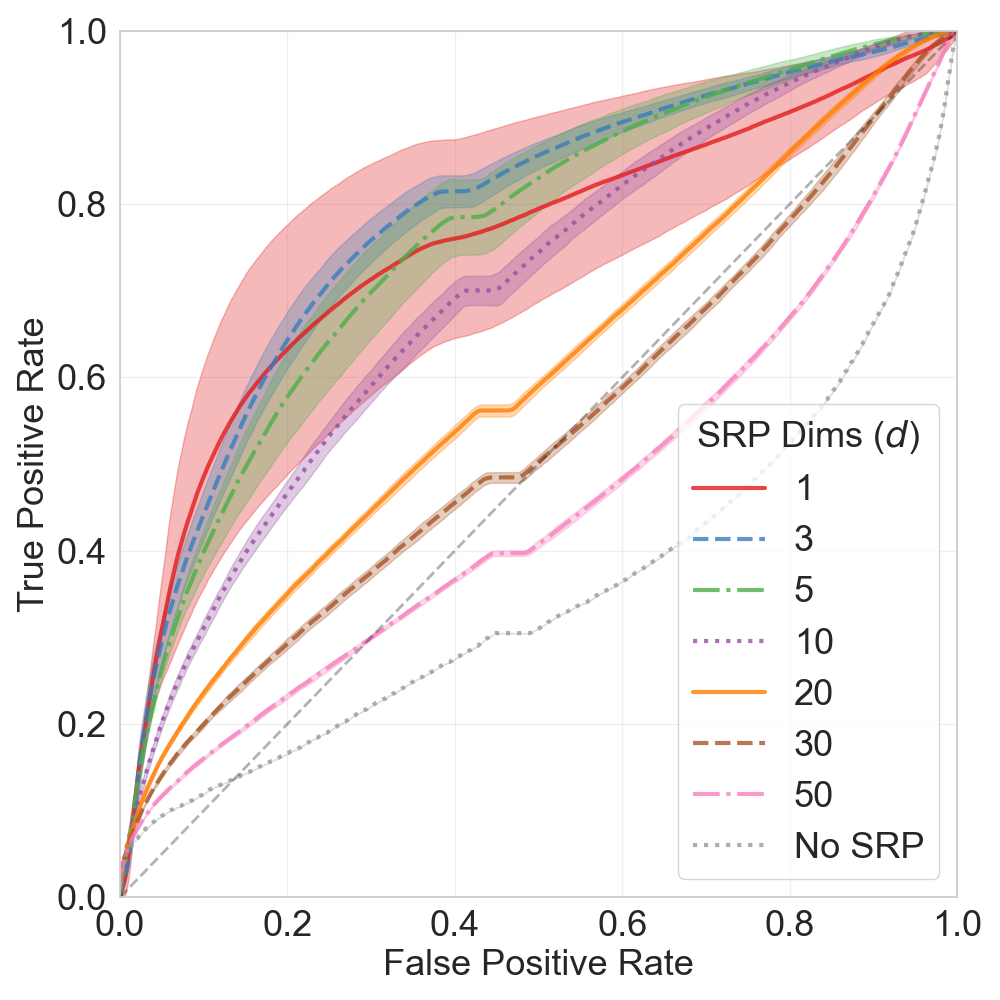}
        \subcaption{Synthetic}
    \end{minipage}%
    \hfill
    \begin{minipage}{0.32\textwidth}
        \centering
        \includegraphics[width=\textwidth]{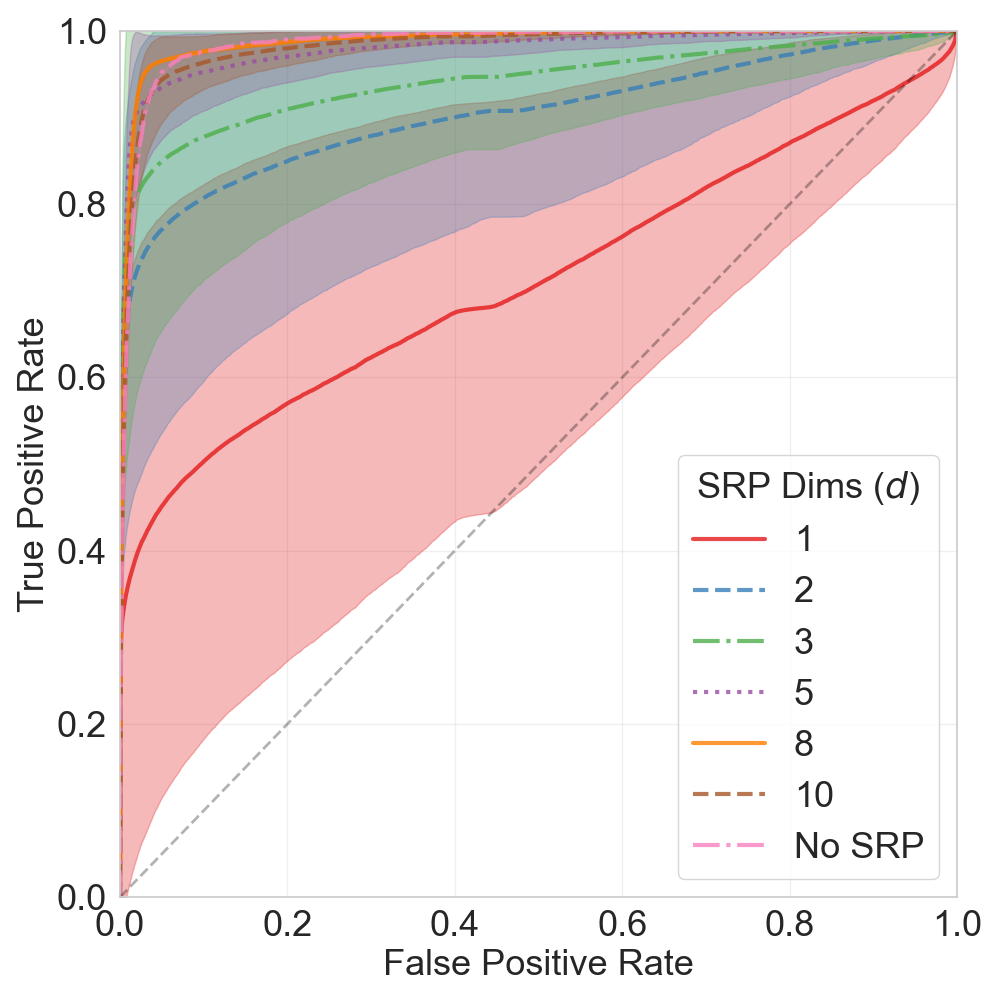}
        \subcaption{SNP}
    \end{minipage}
    \caption{Average ROC curves for ERX with different dimensions using sparse random projection. Shaded areas indicate the standard deviation over ten iterations.}
    \label{roc:projection}
\end{figure*}

Figure \ref{roc:projection} shows the ROC curves of ERX with varying projected dimensions for each dataset, and the detection performance and speed results of the synthetic dataset are shown in Table \ref{tab:projection_synthetic}. 

\begin{table}[h!]
\centering
\caption{Ablation Study - Varying Projected Dimensions ($d$) on the Synthetic Dataset.}
\begin{tabular}{@{}ccc@{}}
\toprule
$d$    & AUC $\pm$ SD      & Speed (LPS) \\ \midrule
$1$    & $0.750 \pm 0.085$      & $\bm{610}$  \\
$3$    & $\bm{0.783 \pm 0.014}$ & $580$       \\
$5$    & $0.759 \pm 0.027$      & $566$       \\
$10$   & $0.698 \pm 0.012$      & $525$       \\
$20$   & $0.596 \pm 0.004$      & $446$       \\
$30$   & $0.532 \pm 0.005$      & $379$       \\
$50$   & $0.452 \pm 0.002$      & $287$       \\
No SRP & $0.356 \pm 0.000$      & $243$       \\ \bottomrule
\end{tabular}
\label{tab:projection_synthetic}
\end{table}

Increasing the number of projected dimensions ($d$) reduces the variance of ERX as more spectral information is retained from the original image after SRP. For the beach and synthetic datasets, originally with 108 and 90 bands, most of the information is retained after 10 dimensions. Balancing the number of projected dimensions is crucial, as it also reduces the processing speed exponentially (Table \ref{tab:projection_synthetic}). ERX is already a fast algorithm due to the quick operation of exponentially moving averages, yet removing SRP altogether clearly shows that it substantially enhances the speed further (by up to 2.5 times on the synthetic dataset, as an example).

However, excessively increasing dimensionality also reduces the detection performance on these datasets. This is seen in the deterioration in the ROC curves and the mean AUC when $d >\approx 5$. This can be expected because of the high correlation between hyperspectral bands and the curse of dimensionality, where too many features hinder anomaly-background separation. On the synthetic dataset especially, ERX becomes worse than a random guessing model after more than 30 dimensions. This suggests that anomalies and background pixels share similar values across many bands, leading to significant redundancy. AUCs below 0.5 indicate that the algorithm is modelling anomalies as the background, which is exacerbated by the high frequency of the anomalies. Reducing dimensionality via SRP mitigates this issue.

Unlike the hyperspectral datasets, the SNP dataset shows much larger standard deviations at low band numbers, even though it originally contains only 13 bands. Detection performance consistently improves and stabilises with increased dimensions instead of degrading after 5 or 10. This is a result of the sparsity condition within SRP, whereby only $\frac{1}{\sqrt{b}}$ of the weights in the randomly generated projection matrix are actually active. Given three projected dimensions ($d=3$), ERX has around 28 and 31 active weights for synthetic and beach datasets respectively, while the SNP dataset only has around 18. Thus, using the recommended density of $\frac{1}{\sqrt{b}}$ for low-dimensional inputs risks losing core information as some key bands may not contribute at all. For lower-dimensional, less correlated data, removing SRP may be more beneficial if speed advantages do not outweigh the loss of information. Nonetheless, for high-dimensional hyperspectral data, sparse random projection to less than 10 dimensions has retained most of the spectral band information, while substantially improving the speed and detection performance of ERX.

\subsubsection{Momentum}

\begin{figure*}[t]
    \centering
    \begin{minipage}{0.45\textwidth}
        \centering
        \includegraphics[width=\textwidth]{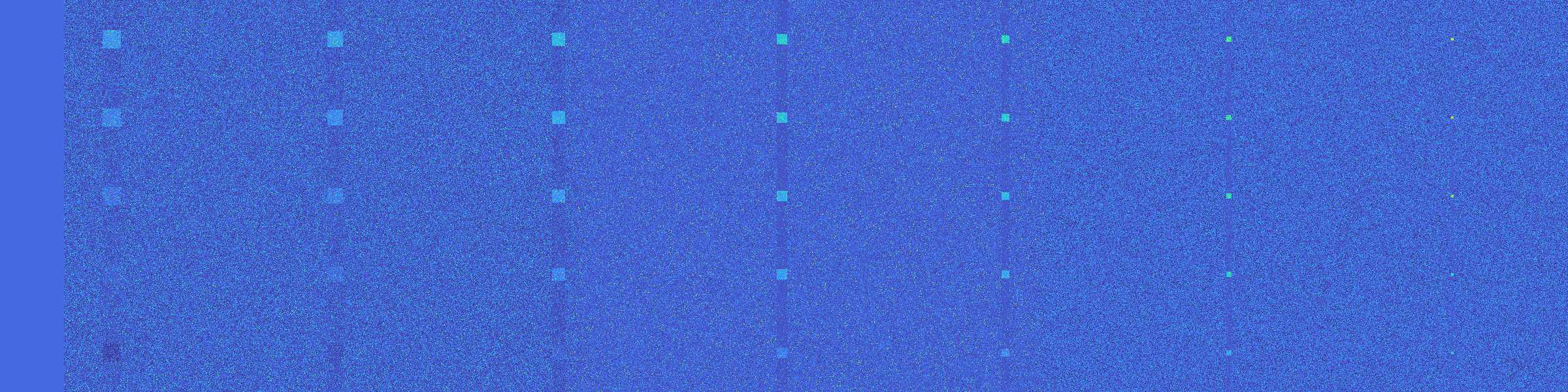}
        \subcaption{$\alpha = 1$}
    \end{minipage}%
    \hfill
    \begin{minipage}{0.45\textwidth}
        \centering
        \includegraphics[width=\textwidth]{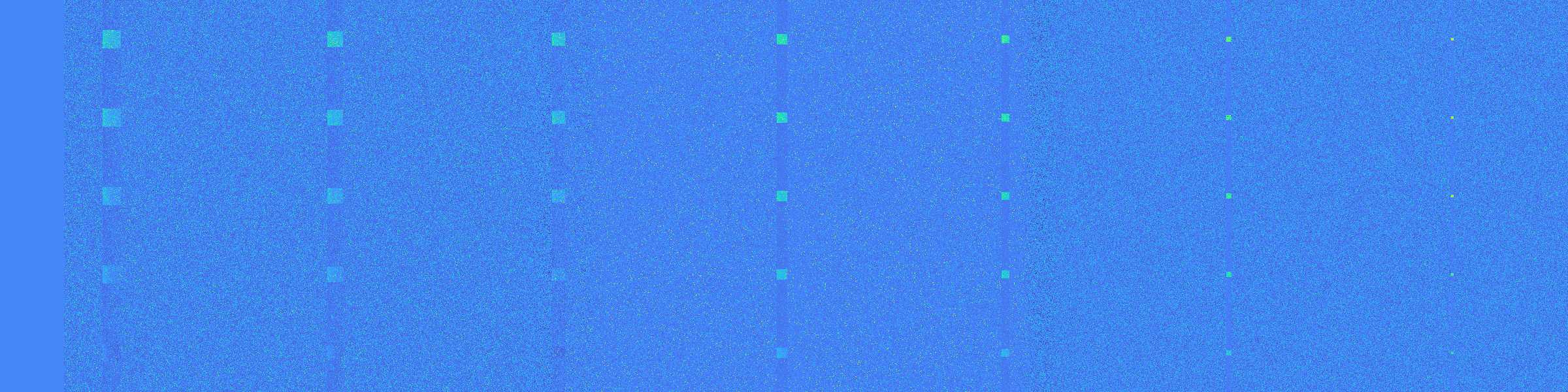}
        \subcaption{$\alpha = 0.1$}
    \end{minipage}% 

    \begin{minipage}{0.45\textwidth}
        \centering
        \includegraphics[width=\textwidth]{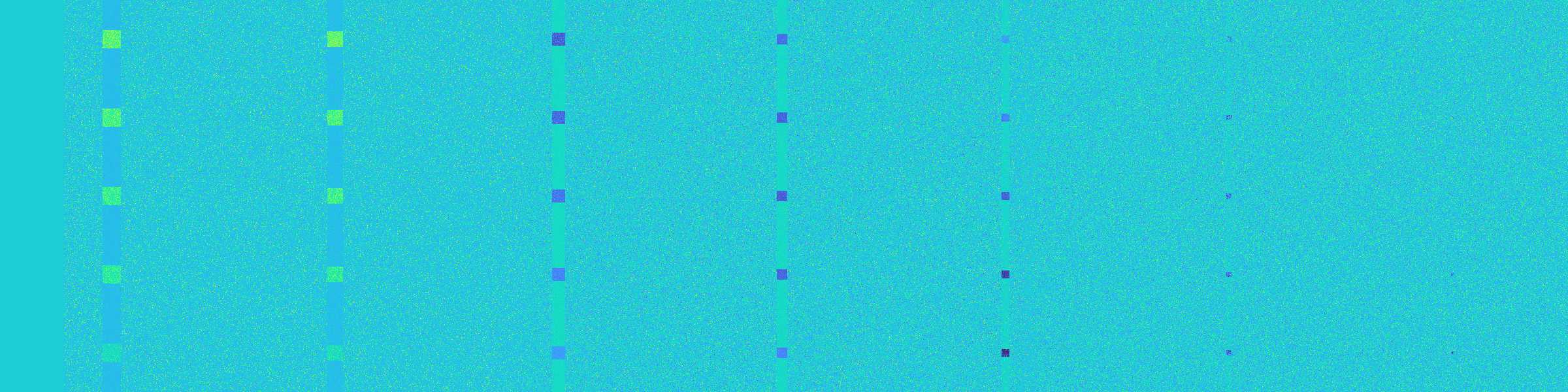}
        \subcaption{$\alpha = 1e-3$}
        \label{fig:momentum_1e-3}
    \end{minipage}%
    \hfill
    \begin{minipage}{0.45\textwidth}
        \centering
        \includegraphics[width=\textwidth]{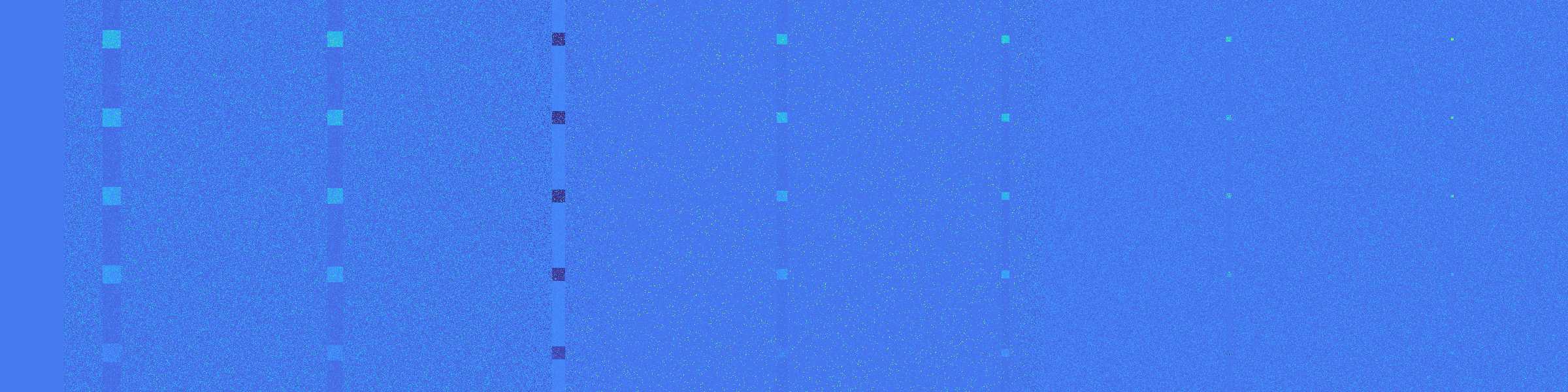}
        \subcaption{No EMAs}
        \label{fig:momentum_imc}
    \end{minipage}% 

    \vspace{0.2cm}
    \begin{minipage}{0.55\textwidth}
        \centering
        \includegraphics[width=\textwidth]{Images/Results/detection/turbo_colourbar.png}
    \end{minipage}% 
    
    \caption{ERX anomaly heatmaps of the synthetic dataset with varying momentum ($\alpha$). ``No EMAs" replaces ERX's exponentially moving averages with incremental mean and covariance updates, equally weighing all previous lines.}
    \label{fig:momentum_maps}
\end{figure*}

The detection results for varying momentum across all datasets are given in Table \ref{tab:momentum_AUC}, and Table \ref{tab:momentum_speed} compares the average speeds when using exponential moving averages (EMAs) and when they are removed. Anomaly heatmaps of the synthetic dataset in Figure \ref{fig:momentum_maps} visually demonstrate how momentum affects ERX's detection performance. 

\begin{table}[h!]
\centering
\caption{Ablation Study - Speed (Lines per Second) With and Without Momentum ($\alpha$).}
\begin{tabular}{@{}cccc@{}}
\toprule
Model & Beach & Synthetic & SNP   \\ \midrule
No EMAs   & $527$ & $535$     & $574$ \\
EMAs  & $\bm{573}$ & $\bm{548}$     & $\bm{599}$ \\ \bottomrule
\end{tabular}
\label{tab:momentum_speed}
\end{table}

\begin{table*}[t]
\centering
\caption{Ablation Study - Mean AUC $\pm$ SD with Varying Momentum ($\alpha$).}
\begin{tabular}{@{}ccccccc@{}}
\toprule
$\alpha$ & Beach                  & Beach (Flipped)        & Synthetic              & Synthetic (Flipped)    & SNP               & SNP (Flipped)     \\ \midrule
$1$      & $0.933 \pm 0.006$      & $0.931 \pm 0.003$      & $0.725 \pm 0.020$      & $0.700 \pm 0.035$      & $0.977 \pm 0.021$ & $0.952 \pm 0.055$ \\
$0.9$    & $0.931 \pm 0.004$      & $0.931 \pm 0.004$      & $0.725 \pm 0.024$      & $0.704 \pm 0.027$      & $0.964 \pm 0.054$ & $0.974 \pm 0.021$ \\
$0.5$    & $0.935 \pm 0.004$      & $0.933 \pm 0.006$      & $0.732 \pm 0.034$      & $0.700 \pm 0.034$      & $0.956 \pm 0.061$ & $0.964 \pm 0.046$ \\
$0.1$    & $0.937 \pm 0.007$      & $0.941 \pm 0.004$      & $\bm{0.773 \pm 0.013}$ & $0.777 \pm 0.029$      & $0.989 \pm 0.001$ & $\bm{0.980 \pm 0.017}$ \\
$0.01$   & $\bm{0.956 \pm 0.005}$ & $\bm{0.956 \pm 0.005}$ & $0.735 \pm 0.023$      & $\bm{0.788 \pm 0.016}$ & $0.979 \pm 0.022$ & $0.925 \pm 0.099$ \\
$1e-3$   & $0.845 \pm 0.045$      & $0.845 \pm 0.034$      & $0.551 \pm 0.013$      & $0.267 \pm 0.034$      & $0.959 \pm 0.093$ & $0.959 \pm 0.078$ \\
$1e-4$   & $0.779 \pm 0.027$      & $0.746 \pm 0.015$      & $0.552 \pm 0.017$      & $0.330 \pm 0.071$      & $0.982 \pm 0.015$ & $0.974 \pm 0.024$ \\
No EMAs      & $0.922 \pm 0.025$      & $0.845 \pm 0.033$      & $0.720 \pm 0.020$      & $0.623 \pm 0.102$      & $\bm{0.990 \pm 0.003}$ & $0.978 \pm 0.026$ \\ \bottomrule
\end{tabular}
\label{tab:momentum_AUC}
\end{table*}

Adjusting the momentum of ERX demonstrates that it improves the detection performance in various scenarios (Table \ref{tab:momentum_AUC}). ERX works well with high momentum on the beach and synthetic datasets, but performance declines after $\alpha < 0.01$. This suggests that when $\alpha \leq 1e-3$, ERX is slow to adapt and over-emphasises past pixels. Small $\alpha$ values overly bias the model toward initial background estimates given by the mean and covariance of the first line, further slowing the adaptation to changing scenery. The synthetic dataset and the flipped version are most affected, as the high anomaly rate further skews the background mean and covariance estimates. Figure \ref{fig:momentum_1e-3} illustrates this, with anomalies blending into the background after the first two columns of anomalous targets. ERX performs consistently across different momentum values on the SNP dataset, as expected due to more uniform scenery and infrequent anomalies. Momentum values of 0.1 and 0.01 generally perform best across all datasets, but it is clear that adjusting the momentum can enhance anomaly detection based on the size and frequency of the anomalies in the scene.

Removing the exponentially moving averages and switching to the incremental mean and covariance (IMC) shows comparable performance in some instances, mostly in the less diverse SNP dataset. Figure \ref{fig:momentum_imc} shows that the IMC still suppresses the background well but misses the third column of anomalies. This anomaly column is located where the background of transitions from vegetation to sand, highlighting an inability to adapt without EMAs. However, a notable difference from removing the EMAs is the reduced mean AUC when the beach and synthetic datasets are flipped. The equal-weighting approach of IMC clearly depends on the starting position, while EMA is more flexible and delivers consistent results. Overall, the exponentially moving averages are a vital part of ERX, and are crucial for adapting to changing scenery and outperforming equal-weight methods in speed and detection.

\subsection{Limitations \& Future Work} 

Creating a Python repository improves accessibility and reproducibility for future projects, but at the cost of potentially slower processing speeds. Implementing these algorithms in faster languages like C++ could improve their speed. With faster processing, more algorithm types, such as those that use kernel and window methods, may become feasible for real-time applications. Performing field tests for extended periods, beyond simulations, is also crucial for validating the practicality of the ERX algorithm.

Despite ERX's overall success, the challenges in separating anomalies from the background in complex datasets remain a persistent issue in pixel-level detection. Grouping pixels into anomalous objects and locating them is a more useful task in practice and may yield better results with object-based metrics. Future research could investigate unsupervised, self-supervised, or synthetically trained object detection algorithms (e.g., YOLO \cite{redmon2016you}) to detect anomalies in any scene containing geometric distortions. Rather than line-by-line detection, using batches of hyperspectral lines could be more practical for detecting and locating anomalous objects in real time, or at least near real time. 

Lastly, selecting optimal thresholds is a key challenge and requires further study for these methods to be fully unsupervised and out-of-the-box anomaly detection algorithms. Without prior knowledge, it is difficult to justify an effective Mahalanobis distance threshold value without data ``snooping" into test datasets. Adaptive threshold selection has been briefly explored \cite{diaz2019line, lin2022real}, leaving room for future contributions. Less supervised models, such as those using few-shot learning or transfer learning with existing hyperspectral libraries, could also improve anomaly detection confidence and reduce the need for optimal threshold selection.

\section{Conclusion} \label{sec:conclusion}

This research introduced the Exponentially moving RX (ERX) algorithm for real-time anomaly detection in hyperspectral line scanning. ERX is fast and scalable to the high dimensions of hyperspectral imagery. It adapts quickly to changing scenery and is robust in detecting anomalies using HSIs that have not been geometrically corrected. Three larger and more complex datasets were introduced to better reflect the challenges of hyperspectral anomaly detection for line scanning, and four other algorithms were implemented onboard a Jetson Xavier NX compute module to evaluate ERX's performance with limited compute power. 

ERX excelled in detection and speed on the beach and synthetic datasets, being the only algorithm fast enough to run in real-time. ERX was the second fastest algorithm on the SNP dataset, and still maintained competitive performance. Sparse random projection effectively reduced dimensionality, enhancing both speed and detection. Maximum benefits are obtained at low dimensions, but excessively low dimensions increase model variance due to inherent randomness. Adjusting ERX's momentum parameter enhances detection performance according to anomaly frequency and scenery variation, while the exponentially moving mean and covariance ensure adaptability and faster processing speeds.

ERX performed best overall, yet the datasets were challenging and open up some promising pathways for future work. These opportunities include field tests to validate algorithm practicality, adaptive and automatic threshold selection, and unsupervised and self-supervised algorithms that group anomalous pixels into objects and locate them.

\section*{Acknowledgments}

The authors thank Dr. Yiwei Mao for his comments on the paper and advice regarding the beach dataset, and Mr. Anthony Quach and Dr. Robert Winter for their general support.

\ifCLASSOPTIONcaptionsoff
  \newpage
\fi



%\bibliographystyle{IEEEtran}
\bibliographystyle{IEEEtran}
\bibliography{bibtex/references}

\end{document}